\documentstyle[12pt,app,twoside,fancyheadings,epsfig]{article}
\newcommand{\be}   {\begin{equation}}
\newcommand{\ee}   {\end  {equation}}
\newcommand{\bea}  {\begin{eqnarray}}
\newcommand{\eea}  {\end  {eqnarray}}

\newcommand{\nim}{Nucl. Inst. and Meth. }

\begin{document}


\thispagestyle{empty}

\begin{center}
  {\large EUROPEAN ORGANIZATION FOR NUCLEAR RESEARCH (CERN)}
\end{center}

\vspace{1cm}
\begin{flushright}
   CERN-EP/2002 - 068 \\
 \today \\
\end{flushright}
\vspace{1cm}

\begin{center}

  {\Large\textbf{Search for anomalous weak dipole \\ 
                 moments of the $\tau$ lepton}} \\[1.cm]

  {\large The ALEPH Collaboration~\footnote{
     See next pages for the list of authors.}}\\[3.cm]

{\large\textbf{Abstract}} \\ 
\end{center}

The anomalous weak dipole moments
of the $\tau$ lepton are 
measured in a data sample collected by ALEPH
from 1990 to 1995
corresponding to an integrated luminosity of 155~pb$^{-1}$.
Tau leptons produced in the reaction
$e^+ e^- \rightarrow \tau^+ \tau^-$ at energies close to the ${\rm Z}$ mass
are studied using their semileptonic decays to
$\pi$, $\rho$, $a_1 \rightarrow \pi 2\pi^0$ or $a_1 \rightarrow 3 \pi$.
The real and imaginary components of both the anomalous weak 
magnetic dipole moment and the CP-violating anomalous weak electric 
dipole moment, 
$ {\rm Re}\,\mu_{\tau}$, ${\rm Im}\,\mu_{\tau}$, ${\rm Re}\,d_{\tau}$ and ${\rm Im}\,d_{\tau}$,
are measured simultaneously by means of a
likelihood fit built from the
full differential cross section. 
No evidence of new physics is found. The following bounds are obtained 
(95\% CL):
$|{\rm Re}\, \mu_{\tau} | < 1.14 \times 10^{-3}$,   
$|{\rm Im}\, \mu_{\tau} | < 2.65 \times 10^{-3}$,
$|{\rm Re}\, d_{\tau} | < 0.91 \times 10^{-3}$, and  
$|{\rm Im}\, d_{\tau} | < 2.01 \times 10^{-3}$. 
\vspace{1.cm}

\begin{center}
  \textit{To be submitted to The European Physical Journal C}
\end{center}

\pagestyle{empty}
\newpage
\small
%
\newlength{\saveparskip}
\newlength{\savetextheight}
\newlength{\savetopmargin}
\newlength{\savetextwidth}
\newlength{\saveoddsidemargin}
\newlength{\savetopsep}
\setlength{\saveparskip}{\parskip}
\setlength{\savetextheight}{\textheight}
\setlength{\savetopmargin}{\topmargin}
\setlength{\savetextwidth}{\textwidth}
\setlength{\saveoddsidemargin}{\oddsidemargin}
\setlength{\savetopsep}{\topsep}
%
%
\setlength{\parskip}{0.0cm}
\setlength{\textheight}{25.0cm}
\setlength{\topmargin}{-1.5cm}
\setlength{\textwidth}{16 cm}
\setlength{\oddsidemargin}{-0.0cm}
\setlength{\topsep}{1mm}
\pretolerance=10000
\centerline{\large\bf The ALEPH Collaboration}
\footnotesize
\vspace{0.5cm}
{\raggedbottom
\begin{sloppypar}
\samepage\noindent
A.~Heister,
S.~Schael
\nopagebreak
\begin{center}
\parbox{15.5cm}{\sl\samepage
Physikalisches Institut das RWTH-Aachen, D-52056 Aachen, Germany}
\end{center}\end{sloppypar}
\vspace{2mm}
\begin{sloppypar}
\noindent
R.~Barate,
I.~De~Bonis,
D.~Decamp,
C.~Goy,
\mbox{J.-P.~Lees},
E.~Merle,
\mbox{M.-N.~Minard},
B.~Pietrzyk
\nopagebreak
\begin{center}
\parbox{15.5cm}{\sl\samepage
Laboratoire de Physique des Particules (LAPP), IN$^{2}$P$^{3}$-CNRS,
F-74019 Annecy-le-Vieux Cedex, France}
\end{center}\end{sloppypar}
\vspace{2mm}
\begin{sloppypar}
\noindent
S.~Bravo,
M.P.~Casado,
M.~Chmeissani,
J.M.~Crespo,
E.~Fernandez,
\mbox{M.~Fernandez-Bosman},
Ll.~Garrido,$^{15}$
M.~Martinez,
A.~Pacheco,
H.~Ruiz
\nopagebreak
\begin{center}
\parbox{15.5cm}{\sl\samepage
Institut de F\'{i}sica d'Altes Energies, Universitat Aut\`{o}noma
de Barcelona, E-08193 Bellaterra (Barcelona), Spain$^{7}$}
\end{center}\end{sloppypar}
\vspace{2mm}
\begin{sloppypar}
\noindent
A.~Colaleo,
D.~Creanza,
M.~de~Palma,
G.~Iaselli,
G.~Maggi,
M.~Maggi,
S.~Nuzzo,
A.~Ranieri,
G.~Raso,$^{23}$
F.~Ruggieri,
G.~Selvaggi,
L.~Silvestris,
P.~Tempesta,
A.~Tricomi,$^{3}$
G.~Zito
\nopagebreak
\begin{center}
\parbox{15.5cm}{\sl\samepage
Dipartimento di Fisica, INFN Sezione di Bari, I-70126
Bari, Italy}
\end{center}\end{sloppypar}
\vspace{2mm}
\begin{sloppypar}
\noindent
X.~Huang,
J.~Lin,
Q. Ouyang,
T.~Wang,
Y.~Xie,
R.~Xu,
S.~Xue,
J.~Zhang,
L.~Zhang,
W.~Zhao
\nopagebreak
\begin{center}
\parbox{15.5cm}{\sl\samepage
Institute of High Energy Physics, Academia Sinica, Beijing, The People's
Republic of China$^{8}$}
\end{center}\end{sloppypar}
\vspace{2mm}
\begin{sloppypar}
\noindent
D.~Abbaneo,
P.~Azzurri,
O.~Buchm\"uller,$^{25}$
M.~Cattaneo,
F.~Cerutti,
B.~Clerbaux,$^{28}$
H.~Drevermann,
R.W.~Forty,
M.~Frank,
F.~Gianotti,
J.B.~Hansen,
J.~Harvey,
D.E.~Hutchcroft,
P.~Janot,
B.~Jost,
M.~Kado,$^{27}$
P.~Mato,
A.~Moutoussi,
F.~Ranjard,
L.~Rolandi,
D.~Schlatter,
O.~Schneider,$^{2}$
G.~Sguazzoni,
W.~Tejessy,
F.~Teubert,
A.~Valassi,
I.~Videau,
J.~Ward
\nopagebreak
\begin{center}
\parbox{15.5cm}{\sl\samepage
European Laboratory for Particle Physics (CERN), CH-1211 Geneva 23,
Switzerland}
\end{center}\end{sloppypar}
\vspace{2mm}
\begin{sloppypar}
\noindent
F.~Badaud,
A.~Falvard,$^{22}$
P.~Gay,
P.~Henrard,
J.~Jousset,
B.~Michel,
S.~Monteil,
\mbox{J-C.~Montret},
D.~Pallin,
P.~Perret
\nopagebreak
\begin{center}
\parbox{15.5cm}{\sl\samepage
Laboratoire de Physique Corpusculaire, Universit\'e Blaise Pascal,
IN$^{2}$P$^{3}$-CNRS, Clermont-Ferrand, F-63177 Aubi\`{e}re, France}
\end{center}\end{sloppypar}
\vspace{2mm}
\begin{sloppypar}
\noindent
J.D.~Hansen,
J.R.~Hansen,
P.H.~Hansen,
B.S.~Nilsson
\begin{center}
\parbox{15.5cm}{\sl\samepage
Niels Bohr Institute, DK-2100 Copenhagen, Denmark$^{9}$}
\end{center}\end{sloppypar}
\vspace{2mm}
\begin{sloppypar}
\noindent
A.~Kyriakis,
C.~Markou,
E.~Simopoulou,
A.~Vayaki,
K.~Zachariadou
\nopagebreak
\begin{center}
\parbox{15.5cm}{\sl\samepage
Nuclear Research Center Demokritos (NRCD), GR-15310 Attiki, Greece}
\end{center}\end{sloppypar}
\vspace{2mm}
\begin{sloppypar}
\noindent
A.~Blondel,$^{12}$
G.~Bonneaud,
\mbox{J.-C.~Brient},
A.~Roug\'{e},
M.~Rumpf,
M.~Swynghedauw,
M.~Verderi,
\linebreak
H.~Videau
\nopagebreak
\begin{center}
\parbox{15.5cm}{\sl\samepage
Laboratoire Leprince-Ringuet, Ecole
Polytechnique, IN$^{2}$P$^{3}$-CNRS, \mbox{F-91128} Palaiseau Cedex, France}
\end{center}\end{sloppypar}
\vspace{2mm}
\begin{sloppypar}
\noindent
V.~Ciulli,
E.~Focardi,
G.~Parrini
\nopagebreak
\begin{center}
\parbox{15.5cm}{\sl\samepage
Dipartimento di Fisica, Universit\`a di Firenze, INFN Sezione di Firenze,
I-50125 Firenze, Italy}
\end{center}\end{sloppypar}
\vspace{2mm}
\begin{sloppypar}
\noindent
A.~Antonelli,
M.~Antonelli,
G.~Bencivenni,
F.~Bossi,
G.~Capon,
V.~Chiarella,
P.~Laurelli,
G.~Mannocchi,$^{5}$
G.P.~Murtas,
L.~Passalacqua,
M.~Pepe-Altarelli$^{4}$
\nopagebreak
\begin{center}
\parbox{15.5cm}{\sl\samepage
Laboratori Nazionali dell'INFN (LNF-INFN), I-00044 Frascati, Italy}
\end{center}\end{sloppypar}
\vspace{2mm}
\begin{sloppypar}
\noindent
J.G.~Lynch,
P.~Negus,
V.~O'Shea,
C.~Raine,$^{6}$
A.S.~Thompson
\nopagebreak
\begin{center}
\parbox{15.5cm}{\sl\samepage
Department of Physics and Astronomy, University of Glasgow, Glasgow G12
8QQ,United Kingdom$^{10}$}
\end{center}\end{sloppypar}
\vspace{2mm}
\begin{sloppypar}
\noindent
S.~Wasserbaech
\nopagebreak
\begin{center}
\parbox{15.5cm}{\sl\samepage
Department of Physics, Haverford College, Haverford, PA 19041-1392, U.S.A.}
\end{center}\end{sloppypar}
\vspace{2mm}
\begin{sloppypar}
\noindent
R.~Cavanaugh,$^{21}$
C.~Geweniger,
P.~Hanke,
V.~Hepp,
E.E.~Kluge,
A.~Putzer,
H.~Stenzel,
K.~Tittel,
M.~Wunsch$^{19}$
\nopagebreak
\begin{center}
\parbox{15.5cm}{\sl\samepage
Kirchhoff-Institut f\"ur Physik, Universit\"at Heidelberg, D-69120
Heidelberg, Germany$^{16}$}
\end{center}\end{sloppypar}
\vspace{2mm}
\begin{sloppypar}
\noindent
R.~Beuselinck,
D.M.~Binnie,
W.~Cameron,
P.J.~Dornan,
M.~Girone,$^{1}$
N.~Marinelli,
J.K.~Sedgbeer,
J.C.~Thompson$^{14}$
\nopagebreak
\begin{center}
\parbox{15.5cm}{\sl\samepage
Department of Physics, Imperial College, London SW7 2BZ,
United Kingdom$^{10}$}
\end{center}\end{sloppypar}
\vspace{2mm}
\begin{sloppypar}
\noindent
V.M.~Ghete,
P.~Girtler,
E.~Kneringer,
D.~Kuhn,
G.~Rudolph
\nopagebreak
\begin{center}
\parbox{15.5cm}{\sl\samepage
Institut f\"ur Experimentalphysik, Universit\"at Innsbruck, A-6020
Innsbruck, Austria$^{18}$}
\end{center}\end{sloppypar}
\vspace{2mm}
\begin{sloppypar}
\noindent
E.~Bouhova-Thacker,
C.K.~Bowdery,
A.J.~Finch,
F.~Foster,
G.~Hughes,
R.W.L.~Jones,
M.R.~Pearson,
N.A.~Robertson
\nopagebreak
\begin{center}
\parbox{15.5cm}{\sl\samepage
Department of Physics, University of Lancaster, Lancaster LA1 4YB,
United Kingdom$^{10}$}
\end{center}\end{sloppypar}
\vspace{2mm}
\begin{sloppypar}
\noindent
K.~Jakobs,
K.~Kleinknecht,
B.~Renk,
\mbox{H.-G.~Sander},
H.~Wachsmuth,
C.~Zeitnitz
\nopagebreak
\begin{center}
\parbox{15.5cm}{\sl\samepage
Institut f\"ur Physik, Universit\"at Mainz, D-55099 Mainz, Germany$^{16}$}
\end{center}\end{sloppypar}
\vspace{2mm}
\begin{sloppypar}
\noindent
A.~Bonissent,
P.~Coyle,
O.~Leroy,
P.~Payre,
D.~Rousseau,
M.~Talby
\nopagebreak
\begin{center}
\parbox{15.5cm}{\sl\samepage
Centre de Physique des Particules, Universit\'e de la M\'editerran\'ee,
IN$^{2}$P$^{3}$-CNRS, F-13288 Marseille, France}
\end{center}\end{sloppypar}
\vspace{2mm}
\begin{sloppypar}
\noindent
F.~Ragusa
\nopagebreak
\begin{center}
\parbox{15.5cm}{\sl\samepage
Dipartimento di Fisica, Universit\`a di Milano e INFN Sezione di Milano,
I-20133 Milano, Italy}
\end{center}\end{sloppypar}
\vspace{2mm}
\begin{sloppypar}
\noindent
A.~David,
H.~Dietl,
G.~Ganis,$^{26}$
K.~H\"uttmann,
G.~L\"utjens,
W.~M\"anner,
\mbox{H.-G.~Moser},
R.~Settles,
W.~Wiedenmann,
G.~Wolf
\nopagebreak
\begin{center}
\parbox{15.5cm}{\sl\samepage
Max-Planck-Institut f\"ur Physik, Werner-Heisenberg-Institut,
D-80805 M\"unchen, Germany\footnotemark[16]}
\end{center}\end{sloppypar}
\vspace{2mm}
\begin{sloppypar}
\noindent
J.~Boucrot,
O.~Callot,
M.~Davier,
L.~Duflot,
\mbox{J.-F.~Grivaz},
Ph.~Heusse,
A.~Jacholkowska,$^{24}$
J.~Lefran\c{c}ois,
\mbox{J.-J.~Veillet},
C.~Yuan
\nopagebreak
\begin{center}
\parbox{15.5cm}{\sl\samepage
Laboratoire de l'Acc\'el\'erateur Lin\'eaire, Universit\'e de Paris-Sud,
IN$^{2}$P$^{3}$-CNRS, F-91898 Orsay Cedex, France}
\end{center}\end{sloppypar}
\vspace{2mm}
\begin{sloppypar}
\noindent
G.~Bagliesi,
T.~Boccali,
L.~Fo\`{a},
A.~Giammanco,
A.~Giassi,
F.~Ligabue,
A.~Messineo,
F.~Palla,
G.~Sanguinetti,
A.~Sciab\`a,
R.~Tenchini,$^{1}$
A.~Venturi,$^{1}$
P.G.~Verdini
\samepage
\begin{center}
\parbox{15.5cm}{\sl\samepage
Dipartimento di Fisica dell'Universit\`a, INFN Sezione di Pisa,
e Scuola Normale Superiore, I-56010 Pisa, Italy}
\end{center}\end{sloppypar}
\vspace{2mm}
\begin{sloppypar}
\noindent
G.A.~Blair,
G.~Cowan,
M.G.~Green,
T.~Medcalf,
A.~Misiejuk,
J.A.~Strong,
\mbox{P.~Teixeira-Dias},
\nopagebreak
\begin{center}
\parbox{15.5cm}{\sl\samepage
Department of Physics, Royal Holloway \& Bedford New College,
University of London, Egham, Surrey TW20 OEX, United Kingdom$^{10}$}
\end{center}\end{sloppypar}
\vspace{2mm}
\begin{sloppypar}
\noindent
R.W.~Clifft,
T.R.~Edgecock,
P.R.~Norton,
I.R.~Tomalin
\nopagebreak
\begin{center}
\parbox{15.5cm}{\sl\samepage
Particle Physics Dept., Rutherford Appleton Laboratory,
Chilton, Didcot, Oxon OX11 OQX, United Kingdom$^{10}$}
\end{center}\end{sloppypar}
\vspace{2mm}
\begin{sloppypar}
\noindent
\mbox{B.~Bloch-Devaux},
P.~Colas,
E.~Lan\c{c}on,
\mbox{M.-C.~Lemaire},
E.~Locci,
P.~Perez,
J.~Rander,
\mbox{J.-P.~Schuller},
B.~Vallage
\nopagebreak
\begin{center}
\parbox{15.5cm}{\sl\samepage
CEA, DAPNIA/Service de Physique des Particules,
CE-Saclay, F-91191 Gif-sur-Yvette Cedex, France$^{17}$}
\end{center}\end{sloppypar}
\vspace{2mm}
\begin{sloppypar}
\noindent
N.~Konstantinidis,
A.M.~Litke,
G.~Taylor
\nopagebreak
\begin{center}
\parbox{15.5cm}{\sl\samepage
Institute for Particle Physics, University of California at
Santa Cruz, Santa Cruz, CA 95064, USA$^{13}$}
\end{center}\end{sloppypar}
\vspace{2mm}
\begin{sloppypar}
\noindent
C.N.~Booth,
S.~Cartwright,
F.~Combley,$^{6}$
M.~Lehto,
L.F.~Thompson
\nopagebreak
\begin{center}
\parbox{15.5cm}{\sl\samepage
Department of Physics, University of Sheffield, Sheffield S3 7RH,
United Kingdom$^{10}$}
\end{center}\end{sloppypar}
\vspace{2mm}
\begin{sloppypar}
\noindent
A.~B\"ohrer,
S.~Brandt,
C.~Grupen,
A.~Ngac,
G.~Prange,
\nopagebreak
\begin{center}
\parbox{15.5cm}{\sl\samepage
Fachbereich Physik, Universit\"at Siegen, D-57068 Siegen,
 Germany$^{16}$}
\end{center}\end{sloppypar}
\vspace{2mm}
\begin{sloppypar}
\noindent
G.~Giannini
\nopagebreak
\begin{center}
\parbox{15.5cm}{\sl\samepage
Dipartimento di Fisica, Universit\`a di Trieste e INFN Sezione di Trieste,
I-34127 Trieste, Italy}
\end{center}\end{sloppypar}
\vspace{2mm}
\begin{sloppypar}
\noindent
J.~Rothberg
\nopagebreak
\begin{center}
\parbox{15.5cm}{\sl\samepage
Experimental Elementary Particle Physics, University of Washington, Seattle, 
WA 98195 U.S.A.}
\end{center}\end{sloppypar}
\vspace{2mm}
\begin{sloppypar}
\noindent
S.R.~Armstrong,
K.~Berkelman,
K.~Cranmer,
D.P.S.~Ferguson,
Y.~Gao,$^{20}$
S.~Gonz\'{a}lez,
O.J.~Hayes,
H.~Hu,
S.~Jin,
J.~Kile,
P.A.~McNamara III,
J.~Nielsen,
Y.B.~Pan,
\mbox{J.H.~von~Wimmersperg-Toeller},
W.~Wiedenmann,
J.~Wu,
Sau~Lan~Wu,
X.~Wu,
G.~Zobernig
\nopagebreak
\begin{center}
\parbox{15.5cm}{\sl\samepage
Department of Physics, University of Wisconsin, Madison, WI 53706,
USA$^{11}$}
\end{center}\end{sloppypar}
\vspace{2mm}
\begin{sloppypar}
\noindent
G.~Dissertori
\nopagebreak
\begin{center}
\parbox{15.5cm}{\sl\samepage
Institute for Particle Physics, ETH H\"onggerberg, 8093 Z\"urich, Switzerland.}
\end{center}\end{sloppypar}
}
\footnotetext[1]{Also at CERN, 1211 Geneva 23, Switzerland.}
\footnotetext[2]{Now at Universit\'e de Lausanne, 1015 Lausanne, Switzerland.}
\footnotetext[3]{Also at Dipartimento di Fisica di Catania and INFN Sezione di
 Catania, 95129 Catania, Italy.}
\footnotetext[4]{Now at CERN, 1211 Geneva 23, Switzerland.}
\footnotetext[5]{Also Istituto di Cosmo-Geofisica del C.N.R., Torino,
Italy.}
\footnotetext[6]{Deceased.}
\footnotetext[7]{Supported by CICYT, Spain.}
\footnotetext[8]{Supported by the National Science Foundation of China.}
\footnotetext[9]{Supported by the Danish Natural Science Research Council.}
\footnotetext[10]{Supported by the UK Particle Physics and Astronomy Research
Council.}
\footnotetext[11]{Supported by the US Department of Energy, grant
DE-FG0295-ER40896.}
\footnotetext[12]{Now at Departement de Physique Corpusculaire, Universit\'e de
Gen\`eve, 1211 Gen\`eve 4, Switzerland.}
\footnotetext[13]{Supported by the US Department of Energy,
grant DE-FG03-92ER40689.}
\footnotetext[14]{Supported by the Leverhulme Trust.}
\footnotetext[15]{Permanent address: Universitat de Barcelona, 08208 Barcelona,
Spain.}
\footnotetext[16]{Supported by Bundesministerium f\"ur Bildung
und Forschung, Germany.}
\footnotetext[17]{Supported by the Direction des Sciences de la
Mati\`ere, C.E.A.}
\footnotetext[18]{Supported by the Austrian Ministry for Science and Transport.}
\footnotetext[19]{Now at SAP AG, 69185 Walldorf, Germany.}
\footnotetext[20]{Also at Department of Physics, Tsinghua University, Beijing, The People's Republic of China.}
\footnotetext[21]{Now at University of Florida, Department of Physics, Gainesville, Florida 32611-8440, USA}
\footnotetext[22]{Now at Groupe d'Astroparticules de Montpellier, Universit\'{e} de Montpellier II, 34095, Montpellier, France}
\footnotetext[23]{Also at Dipartimento di Fisica e Tecnologie Relative, Universit\`a di Palermo, Palermo, Italy.}
\footnotetext[24]{Also at Groupe d'Astroparticules de Montpellier, Universit\'{e} de Montpellier II, 34095, Montpellier, France.}
\footnotetext[25]{Now at SLAC, Stanford, CA 94309, U.S.A.}
\footnotetext[26]{Now at INFN Sezione di Roma II, Dipartimento di Fisica, Universit\'a di Roma Tor Vergata, 00133 Roma, Italy.} 
\footnotetext[27]{Now at Fermilab, PO Box 500, MS 352, Batavia, IL 60510, USA.}
\footnotetext[28]{Now at Institut Inter-universitaire des Hautes Energies (IIHE), CP 230, Universit\'{e} Libre de Bruxelles, 1050 Bruxelles, Belgique.}

\setlength{\parskip}{\saveparskip}
\setlength{\textheight}{\savetextheight}
\setlength{\topmargin}{\savetopmargin}
\setlength{\textwidth}{\savetextwidth}
\setlength{\oddsidemargin}{\saveoddsidemargin}
\setlength{\topsep}{\savetopsep}
\normalsize
\newpage
\pagestyle{plain}
\setcounter{page}{1}

\newpage
\setcounter{page}{1}


\section{Introduction}
 \label{intro}

The anomalous weak dipole moments of the $\tau$ lepton are the
tensorial couplings
of the ${\rm Z} \tau^+ \tau^-$ vertex. They are zero to
first order in the Standard Model (SM). Two types of anomalous weak 
dipole moments can be distinguished:
the magnetic term $\mu_{\tau}$ and the CP-violating electric term
$d_{\tau}$. Here, both the real and the imaginary components
of each anomalous weak dipole moment are explored, i.e.\ 
${\rm Re}\,\mu_{\tau}$, ${\rm Im}\,\mu_{\tau}$, ${\rm Re}\,d_{\tau}$ and ${\rm Im}\,d_{\tau}$.
Radiative corrections in the SM provide nonzero predictions for
$\mu_{\tau}$ and $d_{\tau}$~\cite{Bernabeu,Bern_estimate} which are 
below the present experimental sensitivity. This opens the possibility
to look for deviations from the SM.

There have been many searches for the
CP-violating anomalous weak electric dipole moment of the $\tau$ since 
the beginning of 
LEP~[3-5]. In addition, limits on 
the anomalous weak magnetic dipole moment were 
obtained more recently~\cite{dtau_L3}. 

In this analysis, the previous ALEPH result 
on ${\rm Re}\,d_{\tau}$~\cite{dtau_ALEPH} is updated, and 
${\rm Re}\,\mu_{\tau}$, 
${\rm Im}\,\mu_{\tau}$ and ${\rm Im}\,d_{\tau}$
are determined for the first time in ALEPH.
The data sample 
was collected with the ALEPH detector from 1990 to 1995 
at energies around the ${\rm Z}$ resonance and corresponds to
an integrated luminosity of 155~pb$^{-1}$. 
Tau leptons are generated in the reaction $e^+ e^- \rightarrow \tau^+ \tau^-$
at LEP. 
The method to extract the anomalous weak dipole moments is based 
on a maximum likelihood fit to the data taking into account all $\tau$ spin
terms explicitly, including correlations.
This is the first time that the complete differential 
cross section for the production and decay of the $\tau$ leptons
is considered to estimate the $\tau$ anomalous weak dipole moments. 
The most important semileptonic decays are used: 
$\pi$, $\rho$, $a_1 \rightarrow \pi 2\pi^0$ and $a_1 \rightarrow 3 \pi$.
The $\tau$ spin information is recovered using optimal polarimeters 
which are different for each decay.
The selection and particle identification make use of tools already
developed in previous analyses ~[6-9]. 

The text is organized as follows. The theoretical framework is
introduced in Section~\ref{theory}.
The most important ALEPH subdetectors for this analysis
are covered in Section~3.
The data analysis procedure is explained in
Section~\ref{analysis},
emphasizing the new features of the analysis. The more relevant
systematic uncertainties are then discussed in Section~\ref{sys_errors}. 
The results
and conclusions are presented in Section~\ref{results}.

\section{Theoretical framework}
 \label{theory}

\subsection{Production cross section}

The currents assumed for photon and ${\rm Z}$ exchange
in $e^+ e^- \rightarrow \tau^+ \tau^-$ production are 
\bea
\Gamma^{\mu \, (\gamma)}_{f}&=&i Q_f {\rm e} \gamma^{\mu}, \; \; \; \; \; \mbox{with $f$ = $e$, $\tau$} 
 \; , \nonumber \\
\Gamma^{\mu \, ({\rm Z})}_e&=&i{\rm e} \left[ v_e \gamma^{\mu} - a_e \gamma^{\mu}\gamma_5 
\right] \, ,  \nonumber \\
\Gamma^{\mu \, ({\rm Z})}_{\tau}&=&i{\rm e} \left[ v_{\tau} \gamma^{\mu} - a_{\tau} \gamma^{\mu}\gamma_5 
+ i\frac{\mu_{\tau}}{2 m_{\tau}}\sigma^{\mu \nu}q_{\nu}
+ \frac{d_{\tau}}{2 m_{\tau}}\gamma_5 \sigma^{\mu \nu}q_{\nu} 
\right]\;, 
\eea
\noindent
where $Q_f {\rm e}$ is the fermion charge;
$a_e$, $a_{\tau}$, $v_e$ and $v_{\tau}$ are the axial vector and vector 
couplings of the SM;  
$\mu_{\tau}$ and $d_{\tau}$ are the anomalous weak magnetic and anomalous weak 
electric dipole
moments of the $\tau$. In the previous expression both anomalous weak dipole 
moments are dimensionless quantities. However, 
the anomalous weak electric dipole moment is often quoted in the literature 
in units of ${\rm e \, cm}$, by defining the
contribution of this dipole moment to the current as
$i d_{\tau}\gamma_5 \sigma^{\mu \nu}q_{\nu}$. 
These different notations are related by the
conversion factor
e/2$m_{\tau}$ = $5.552 \times 10^{-15} {\rm e \, cm}$.

Using the currents in Eq.~1,
the differential cross section can be expressed as~\cite{Stiegler}
\bea
\frac {d\sigma} {d\cos \theta_{\tau} }(\vec{s}_1,\vec{s}_2) &=& R_{00} +    \sum_{\mu=1,3}   R_{\mu0} s_1^{\mu} 
                                      +    \sum_{{\nu}=1,3}   R_{0{\nu}} s_2^{\nu}
                                      +    \sum_{{\mu},{\nu}=1,3} R_{{\mu}{\nu}} s_1^{\mu} s_2^{\nu} \, .
\label{eq_xsect}
\eea
\noindent
The $R_{\mu \nu}$ terms are functions of the fermion couplings and of the $\tau$
production angle $\theta_{\tau}$; 
$\vec{s}_1$ and $\vec{s}_2$ are unit vectors
chosen as the quantisation axes for the spin measurement of
the $\tau^+$ and the $\tau^-$, respectively,
in their corresponding rest frames.

The following reference frame has been chosen:
the $z$ axis is in
the outgoing $\tau^+$ direction and the incoming $e^+$ is in the $yz$ plane.
The $x$ component is therefore normal to the production plane.
The $y$ component is called transverse.

Several  $R_{\mu \nu}$ terms have been already measured by ALEPH. 
Defining $ \left(  R_{\mu \nu} \right)_{\pm} \equiv (R_{\mu \nu} \pm R_{\nu \mu}) $, 
these terms are the following:
\begin{itemize}
\item[-] $R_{00}$ = $d \sigma$/ $d \cos \theta_{\tau}$~\cite{R_00_reference},
\item[-] $(R_{03})_+/R_{00} = P_{\tau}(\cos \theta_{\tau})$ is the longitudinal
      polarisation of the $\tau$~\cite{aleph_polarisation_paper},
\item[-] $R_{22}/R_{00} = - R_{11}/R_{00}$ are the transverse-transverse
      and normal-normal spin correlations~\cite{ALEPH_corr},
\item[-] $(R_{21})_+/R_{00}$ are the transverse-normal
      spin correlations~\cite{ALEPH_corr}.
\end{itemize}

The $R_{\mu \nu}$ terms most sensitive to 
${\rm Re}\,\mu_{\tau}$, ${\rm Im}\,\mu_{\tau}$, ${\rm Re}\,d_{\tau}$ and ${\rm Im}\,d_{\tau}$
are presented below. 
These terms are obtained from Ref.~\cite{Stiegler} after some algebra.
The SM contributions are separated from the anomalous (anm)
contributions. 

${\rm Re}\,\mu_{\tau}$:

\bea
\left. \left( R_{02} \right)_+ \right|_{\rm SM}  &\propto &   
\frac{2}{\gamma_{\tau}} \sin \theta_{\tau} \, |v_{\tau}|^2 Re(v_e a_e^{*}) +
\frac{1}{\gamma_{\tau}} \sin \theta_{\tau} \, \cos \theta_{\tau} \, (|a_e|^2 + |v_e|^2)Re(v_{\tau}a_{\tau}^*) \nonumber \\
\left. \left( R_{02} \right)_+ \right|_{\rm anm} &\propto & 
   \gamma_{\tau} \sin \theta_{\tau} \, \cos \theta_{\tau} \,( |a_e|^2 + |v_e|^2) Re(a_{\tau} \mu_{\tau}^{*}) \nonumber \\
&&\quad{}+          \frac{2(\gamma_{\tau}^2+1)}{\gamma_{\tau}} \sin \theta_{\tau} \, Re(v_e a_e^{*})Re({v_{\tau}} \mu_{\tau}^{*}) 
+   2 \gamma_{\tau} \sin \theta_{\tau} \, Re(v_e a_e^{*})|\mu_{\tau}|^2 
\eea
\bea
\left. \left( R_{32} \right)_+ \right|_{\rm SM} &\propto &  
\frac{2}{\gamma_{\tau}} \sin \theta_{\tau} \, Re(v_e a_e^{*}) Re(v_{\tau} a_{\tau}^{*}) +
\frac{1}{\gamma_{\tau}} \sin \theta_{\tau} \, \cos \theta_{\tau} \, (|a_e|^2 + |v_e|^2) |v_{\tau}|^2  \nonumber \\
\left. \left( R_{32} \right)_+ \right|_{\rm anm} &\propto & 
   \frac{\gamma_{\tau}^2+1}{\gamma_{\tau}} \sin \theta_{\tau} \, \cos \theta_{\tau} \,( |a_e|^2 + |v_e|^2) Re(v_{\tau} \mu_{\tau}^{*}) \nonumber \\
&&\quad{}+  2 \gamma_{\tau} \sin \theta_{\tau} \, Re(v_e a_e^{*})Re(a_{\tau}{\mu}_{\tau}^{*})
+  \gamma_{\tau} \sin \theta_{\tau} \, \cos \theta_{\tau} \,( |a_e|^2 + |v_e|^2) |\mu_{\tau}|^2  
\eea

${\rm Im}\,\mu_{\tau}$:

\bea
\left. \left( R_{31} \right)_+ \right|_{\rm SM} &\propto &  
\frac{1}{\gamma_{\tau}} \sin \theta_{\tau} \, \cos \theta_{\tau} \,( |a_e|^2 + |v_e|^2 ) 
Im(v_{\tau}^{*} a_{\tau})  \nonumber \\
\left. \left( R_{31} \right)_+ \right|_{\rm anm} &\propto & 
   \gamma_{\tau} \sin \theta_{\tau} \, \cos \theta_{\tau} \,( |a_e|^2 + |v_e|^2) 
Im(a_{\tau} \mu_{\tau}^{*}) \nonumber \\
&&\quad{}+  \frac{2(\gamma_{\tau}^2-1)}{\gamma_{\tau}} \sin \theta_{\tau} \, 
Re(v_e a_e^{*})Im({v_{\tau}} {\mu}_{\tau}^{*})  
\eea
\bea
%
\left. \left( R_{01} \right)_+ \right|_{\rm SM} &\propto &  
\frac{2}{\gamma_{\tau}} \sin \theta_{\tau} \, Re(v_e a_e^{*}) Im(v_{\tau}^{*}a_{\tau})  \nonumber \\
\left. \left( R_{01} \right)_+ \right|_{\rm anm} &\propto & 
   \frac{\gamma_{\tau}^2-1}{\gamma_{\tau}}  \sin \theta_{\tau} \, \cos \theta_{\tau} \,( |a_e|^2 + |v_e|^2) Im(v_{\tau} \mu_{\tau}^{*}) \nonumber \\
&&\quad{}+  2 \gamma_{\tau} \sin \theta_{\tau} \, Re(v_e a_e^{*})Im(a_{\tau}{\mu}_{\tau}^{*}) 
\eea

${\rm Re}\,d_{\tau}$:

%
\bea
\left. \left( R_{01} \right)_- \right|_{\rm SM} &\propto & 0  \nonumber \\
\left. \left( R_{01} \right)_- \right|_{\rm anm} &\propto & -
   \gamma_{\tau} \sin \theta_{\tau} \, \cos \theta_{\tau} \, (|a_e|^2+ |v_e|^2) 
Re(a_{\tau}d_{\tau}^{*})
\nonumber \\
&&\quad{}-    2 \gamma_{\tau} \sin \theta_{\tau} \, Re(v_e a_e^{*})
\left[ Re(v_{\tau} d_{\tau}^{*}) +
Re({\mu}_{\tau} d_{\tau}^{*}) \right]  
\eea
\bea
%
\left. \left( R_{31} \right)_- \right|_{\rm SM} &\propto &  0  \nonumber \\
\left. \left( R_{31} \right)_- \right|_{\rm anm} &\propto &-
   \gamma_{\tau} \sin \theta_{\tau} \, \cos \theta_{\tau} \, (|a_e|^2+ |v_e|^2) \left[ Re(v_{\tau} d_{\tau}^{*}) +
Re({\mu}_{\tau} d_{\tau}^{*}) \right]  \nonumber \\ 
  &&\quad{}- 2 \gamma_{\tau} \sin \theta_{\tau} \, Re(v_e a_e^{*})Re(a_{\tau} d_{\tau}^{*}) 
\eea

${\rm Im}\,d_{\tau}$:

\bea
%
%
%
\left. \left( R_{32} \right)_- \right|_{\rm SM} &\propto &  0  \nonumber \\
\left. \left( R_{32} \right)_- \right|_{\rm anm} &\propto & 
   \gamma_{\tau} \sin \theta_{\tau} \, \cos \theta_{\tau} \, (|a_e|^2+ |v_e|^2) Im(a_{\tau}d_{\tau}^{*}) \nonumber \\ 
&&\quad{}+    2 \gamma_{\tau} \sin \theta_{\tau} \, Re(v_e a_e^{*})
\left[ Im(v_{\tau}d_{\tau}^{*}) +
Im({\mu}_{\tau}d_{\tau}^{*}) \right] 
\eea
\bea
%
%
\left. \left( R_{02} \right)_- \right|_{\rm SM} &\propto &  0   \nonumber \\
\left. \left( R_{02} \right)_- \right|_{\rm anm} &\propto & 
   \gamma_{\tau} \sin \theta_{\tau} \, \cos \theta_{\tau} \, (|a_e|^2+ |v_e|^2)
\left[ Im(v_{\tau}d_{\tau}^{*}) +
Im({\mu}_{\tau}d_{\tau}^{*}) \right]  \nonumber \\
   &&\quad{}+ 2 \gamma_{\tau} \sin \theta_{\tau} \, Re(v_e a_e^{*})Im(a_{\tau}d_{\tau}^{*})
\eea

\noindent
Taking into account that $a_l \gg v_l$, the terms can be ordered in 
sensitivity, and the most sensitive term for each
anomalous weak dipole moment is presented first.
The quantity $\gamma_{\tau}$ is computed as $\sqrt{s}/2 m_{\tau}$.
The photon exchange terms are omitted 
from these expressions for simplicity, although
they are taken into account in the final results. 

The anomalous weak dipole moments are extracted including all 
$R_{\mu \nu}$ terms in
a maximum likelihood fit. In this analysis $ \left( R_{31} \right)_{+}$, 
the most sensitive term to ${\rm Im}\, \mu_{\tau}$, 
is used for the first time as proposed in Ref.~\cite{Imaginary}.
The terms
$ \left( R_{02} \right)_{+}$, $ \left( R_{01} \right)_{-}$
and $ \left( R_{32} \right)_{-}$ 
were previously used
in other measurements of the anomalous weak dipole moments.

\subsection{Tau decay}

For each $\tau$ decay mode, the differential partial width of a polarised 
$\tau$ is written as
\begin{equation}
 d \Gamma (\vec{s}) = W (1+ \vec{h} \cdot \vec{s}) \;  d X \: ,
\label{eq_dif_decay_3}
\end{equation}
\noindent
using the expressions for $W$ and $\vec{h}$ from the
TAUOLA Monte Carlo program~\cite{Tauola};
$W$ is the differential partial width of an unpolarised 
$\tau$, and the 
$\vec{h}$ vector 
is the polarimeter of the particular decay mode considered. Both
$W$ and $\vec{h}$ depend on the four-momenta of
the final state particles in the $\tau$ rest frame, and 
they are different for each decay topology. The 
simplest expressions are those of the $\tau$ decay into $\pi$.
In this case,
$\vec{h}_{\pi}$ is proportional to the $\pi$
momentum in the $\tau$ rest frame and $W_{\pi}$ is a constant.
In the above equation, $X$ is a set 
of independent variables describing the full decay configuration.
The number of elements of the set depends on the number of particles
in the final state. The set $X_{\pi}$ denotes the set of variables
expressing the $\pi$ direction in the $\tau$ rest frame.

In this analysis, the expressions for $W$ and $\vec{h}$ 
allow the spin information for 
all the $\tau$ decays to be recovered optimally. 

\subsection{The full differential cross section}

Once the production cross section and the partial decays are introduced,
the full differential cross section of  
$e^+ e^- \rightarrow \tau^+ \tau^- \rightarrow 
x^+_1 x^-_2 \bar{\nu}_{\tau} \nu_{\tau}$
is built following
Refs.~\cite{Tesis_NRius,Jadach_1984}, namely

\begin{equation}
\frac {d\sigma} {d\cos \theta_{\tau} d X_1 d X_2} = 
                                          4\frac{W_1}{\Gamma_{\tau}} \frac{W_2}{\Gamma_{\tau}} 
                                  \left[        R_{00} +    \sum_{{\mu}=1,3}   R_{{\mu}0} h_1^{\mu} 
                                      +    \sum_{{\nu}=1,3}   R_{0{\nu}} h_2^{\nu}
                                      +    \sum_{{\mu},{\nu}=1,3} R_{{\mu}{\nu}} h_1^{\mu} h_2^{\nu} \right] \, . 
\label{eq_final_xsect}
\end{equation}
\noindent
In this equation $\Gamma_{\tau}$ is the total
$\tau$ width, $X_1$ and $X_2$ are the sets of independent variables,
and $\vec{h}_1$ and $\vec{h}_2$ are the polarimeters for the decay
of the $\tau^+$ and the $\tau^-$, respectively.

\begin{figure}[htbp]
\begin{center}
\mbox{\epsfig{file=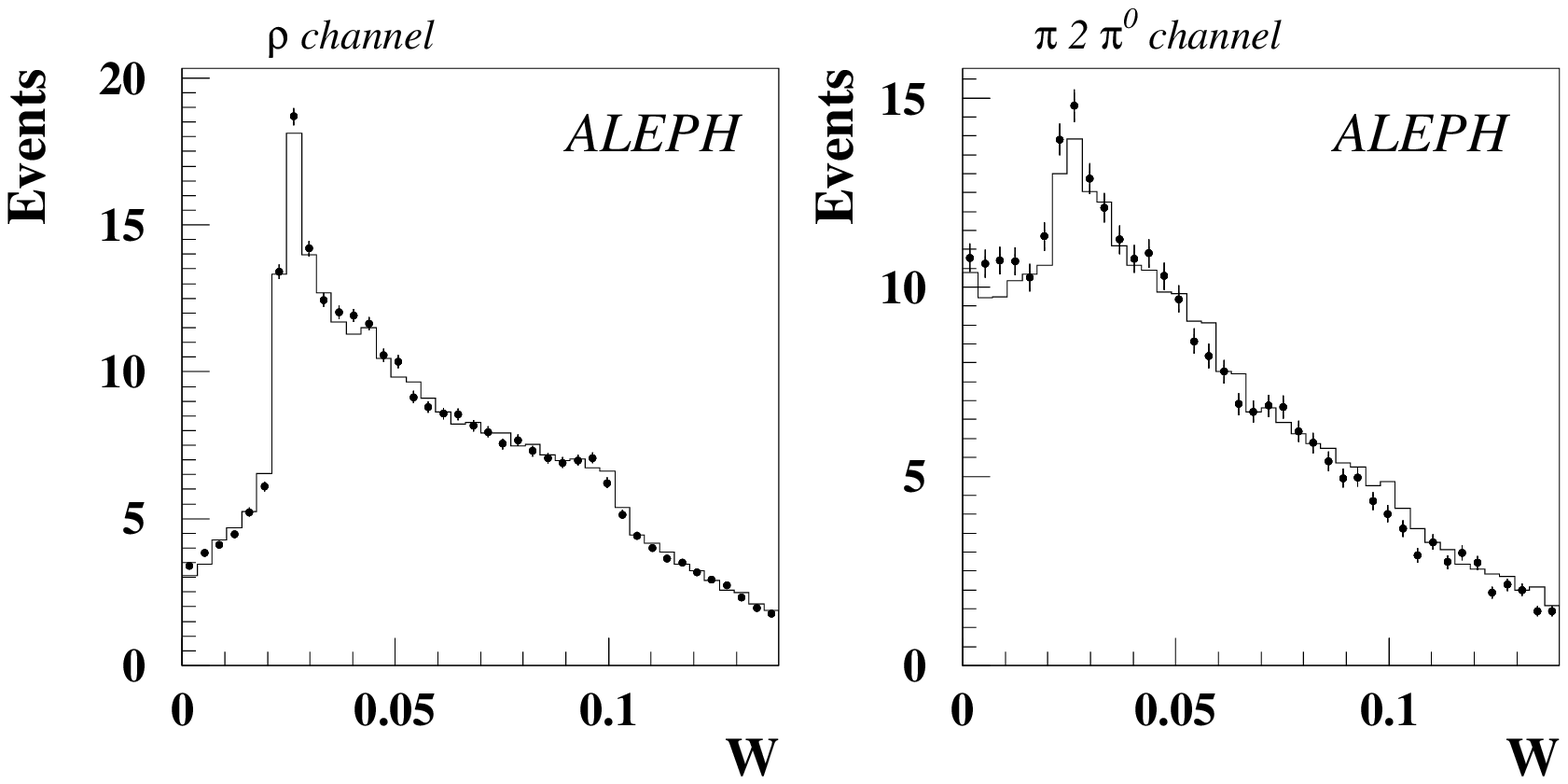,width=15.cm}}
\mbox{\epsfig{file=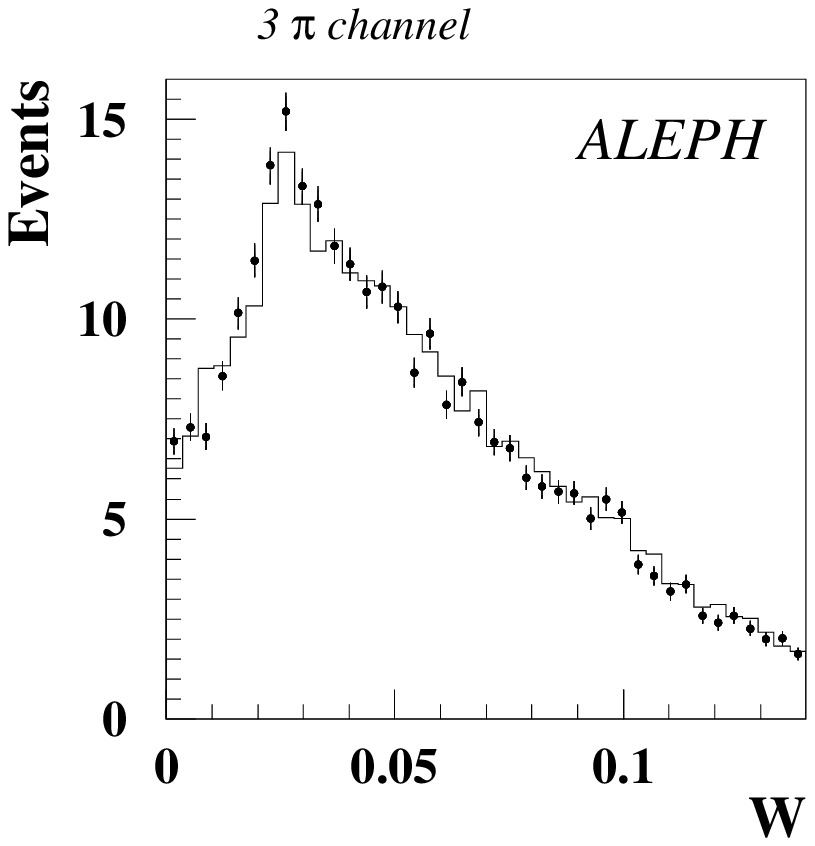,width=9.cm}}
\caption[\protect\footnotesize  Comparison of the 
$W$ distribution for the data and the Monte Carlo simulation
]
{\protect\footnotesize
The $W$ observable for the $\rho$, $\pi 2\pi^0$ and $3 \pi$ decays.
The points are the data and the histogram is the simulation.
Both distributions are normalised to unit area in each plot. Only the statistical errors
are included.
}
\label{f_obs_w}
\end{center}
\end{figure}

\begin{figure}[htbp]
\begin{center}
\mbox{\epsfig{file=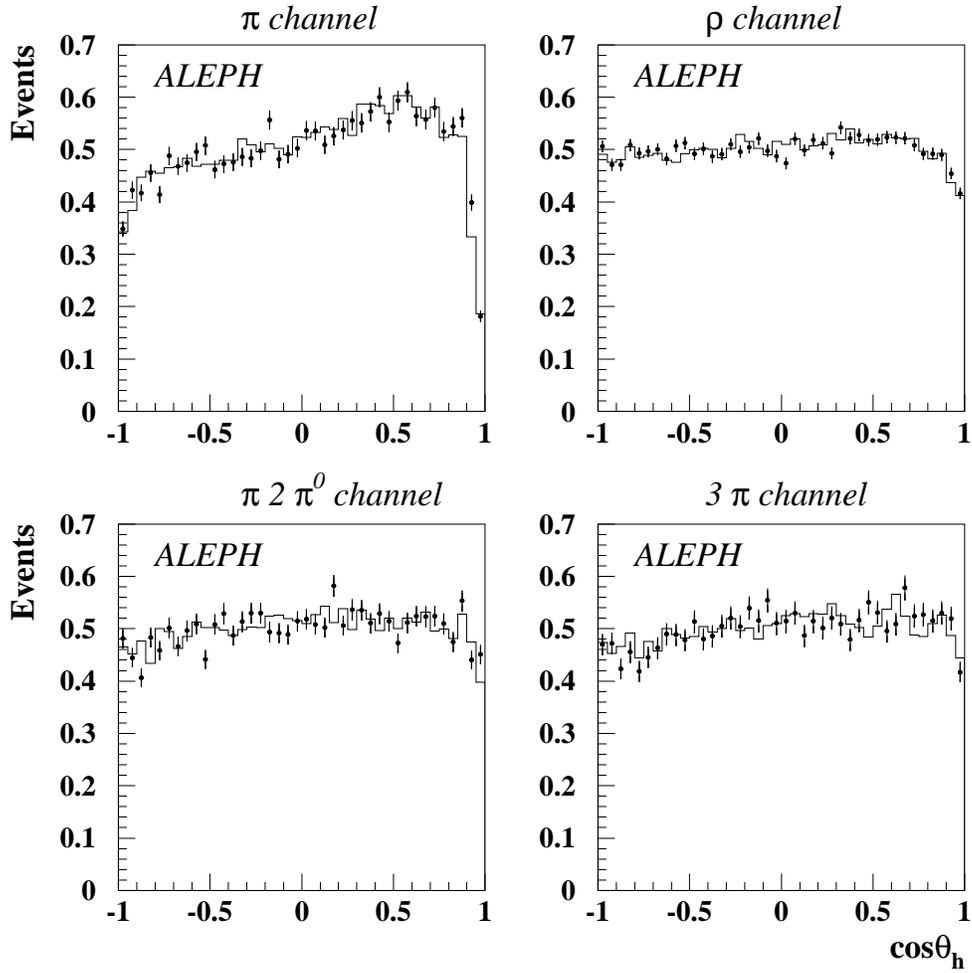,width=15.cm}}
\caption[\protect\footnotesize Comparison of the 
$\cos \theta_h$ distribution for the data and the Monte Carlo simulation
]
{\protect\footnotesize
The $\cos \theta_h$ observable for the four decay topologies.
The points are the data and the histogram is the simulation.
Both distributions are normalised to unit area in each plot. Only the statistical errors
are included.
}
\label{f_obs_cs}
\end{center}
\end{figure}

With the definitions $\bar{R}_{\mu \nu} = R_{\mu \nu}/R_{00}$,
$H^{\mu} = W h^{\mu}/\Gamma_{\tau}$ ($\mu, \nu = 0, \ldots, 3$),
and $h^0 = 1$,
the likelihood of an event with the final state topology $ij$
is written as
\bea\lefteqn{
 L_{ij}(\mu_{\tau},d_{\tau}|\theta_{\tau},W_1,\cos {\theta}_{h_1}, \phi_{h_1},
W_2,\cos {\theta}_{h_2}, \phi_{h_2})  = } \\ \nonumber
&&\rule{30mm}{0pt}
\sum_{\mu,\nu =0,\ldots,3}
\bar{R}_{\mu \nu}(\mu_{\tau},d_{\tau},\theta_{\tau})  
H_i^{\mu}(
W_1,\cos {\theta}_{h_1}, \phi_{h_1})
H_j^{\nu}(
W_2,\cos {\theta}_{h_2}, \phi_{h_2} )
\:.\qquad
\label{eq_like_theory}
\eea
\noindent
The indices ($i$, $j$) refer to the decay mode of each $\tau$, 
with $i, j = \pi, \rho, \pi 2\pi^0, 3\pi$, and
the quantities
($W_1,\cos {\theta}_{h_1}, \phi_{h_1}$) and
($W_2,\cos {\theta}_{h_2}, \phi_{h_2}$) are the
observables related to the decay of the $\tau^+$ and the $\tau^-$, 
respectively.
The angles
(${\theta}_{h_1}$, $\phi_{h_1}$,  
${\theta}_{h_2}$, $\phi_{h_2}$) are the polar and azimuthal angles
of the polarimeters of each $\tau$, in the reference
frame introduced in Section 2.1.
The above likelihood is also a function of the centre-of-mass energy. 
The distributions of the hemisphere observables  
$W$ and $\cos {\theta}_{h}$ are presented in 
Figs.~\ref{f_obs_w} and~\ref{f_obs_cs}.

This likelihood fulfills the normalisation condition 
\begin{equation}
\sum_{ij} \int L_{ij}(\mu_{\tau},d_{\tau}|\theta_{\tau} ,W_1, \cos {\theta}_{h_1}, \phi_{h_1}, 
W_2, \cos {\theta}_{h_2}, \phi_{h_2}   ) 
d{X_1} 
d{X_2}
= 1 \: .
\end{equation}
\noindent
This is
the integral over all possible decay
parameters and all possible decay topologies $ij$ for a given
$e^+ e^- \rightarrow \tau^+ \tau^-$ event. 
The normalisation is such that the likelihood depends only upon the net spin
polarisation of the produced $\tau$ pairs, and not upon
$R_{00}$ = $d \sigma$/$d \cos \theta_{\tau}$.

\section{Apparatus}
\label{aleph}
The ALEPH detector is described in detail in~\cite{ALEPH_1} and its performance
in~\cite{ALEPH_2}.

Charged particles are measured with a high resolution silicon vertex detector (VDET),
a cylindrical drift chamber (ITC), and a large time projection chamber (TPC). The
momentum resolution in the axial magnetic field of 1.5$\,$T provided by a superconducting
solenoid is $\Delta p / p^2 = 0.6 \times 10^{-3}$(GeV/$c$)$^{-1}$ for high 
momentum tracks. The impact parameter resolutions for high momentum
tracks with hits in all three
subdetectors are $\sigma_{r \phi} = 23 \,\mu$m and  $\sigma_z = 28 \,\mu$m. 

The tracking devices are surrounded by the electromagnetic calorimeter (ECAL), which
is a highly segmented lead/proportional-wire-chamber calorimeter. The calorimeter
is read out via cathode pads arranged in projective towers covering 
$0.9^{\circ} \times 0.9^{\circ}$ in solid angle and summing the deposited energy in three sections
of depth. A second readout is provided by the signals from the anode wires. The
energy resolution is $\sigma / E = 0.009 + 0.18/ \sqrt{E{\rm(GeV)}}$.

The ECAL is inside the solenoid, which is followed by the hadron calorimeter (HCAL).
Hadronic showers are sampled by 23 planes of streamer tubes giving a digital hit
pattern and an analog signal on pads, which are also arranged in projective towers.
This calorimeter is used in this analysis
to discriminate between pions and muons. Outside the HCAL
there are two layers of muon chambers providing additional information for $\mu$
identification.

\section{Data analysis}
 \label{analysis}
%
%
\subsection{Selection and decay classification}
\label{selection}

Events from ${\rm Z} \rightarrow \tau^+ \tau^-$ are retained using a global selection
in which each event is divided into two hemispheres along the thrust axis. 
The selection is that used in the ALEPH measurement of 
$P_{\tau}(\cos \theta_{\tau})$ with the $\tau$ direction 
method~\cite{aleph_polarisation_paper}. Additional information can be
found in~\cite{br_had} and references therein. 

The charged particle identification is based on a likelihood method which
assigns a set of probabilities to each particle.
A detailed description of the method can be found in Refs.~\cite{first_taupid,br_lep}. 
The probability set for each particle is obtained from ($i$) the 
specific ionisation dE/dx in the TPC,
($ii$) the longitudinal and transverse shower profiles in ECAL near the
extrapolated track and ($iii$) the energy and average shower width in HCAL,
together with the number out of the last ten of HCAL planes that
fired and the number of hits in the muon chambers.

The photon and $\pi^0$ reconstruction is performed with a likelihood method
which first distinguishes between
genuine and fake photons produced by hadronic
interactions in ECAL or by electromagnetic shower fluctuations~\cite{br_had}.
All photon pairs in each hemisphere are then assigned a probability
of being generated by a $\pi^0$. High energy $\pi^0$ with overlapping
showers are reconstructed through an analysis of the spatial energy deposition
in the ECAL towers. All the remaining single photons are considered
and those with a high probability of being a genuine photon are selected
as $\pi^0$ candidates. Finally, photon conversions are identified
following the procedure described in~\cite{br_had}. They are added to the list
of good photons and are included in the $\pi^0$ reconstruction.

The $\tau$ decay classification depends on the number of charged tracks and
their identification, and on the number of reconstructed $\pi^0$. It follows
the classification for the measurement of $P_{\tau}(\cos \theta_{\tau})$ with 
the $\tau$ direction method, described in~\cite{aleph_polarisation_paper} 
and the references therein. 

The $\tau$ selection efficiencies and the
background fractions for the data, presented in Table~1, are estimated
from the Monte Carlo simulation. Only statistical errors are quoted.
In this data sample the only relevant contamination arises from
$\tau^+ \tau^-$ events with misidentified decay modes ($\tau$ background).
\begin{table}[tbhp]
\caption{\protect\footnotesize
Selection efficiencies and $\tau$
background for the different decay channels,
obtained from the Monte Carlo simulation and
presented with statistical errors only.
For this table all identified events are retained.
}
\label{t_eff_bkg}
\begin{center}
\begin{tabular}{|c|c|c|} \hline
  $\tau$ decay     & Efficiency (\%)  & $\tau$ Background (\%) \\ \hline \hline
  $\pi$-$\pi$         & 57.57 $\pm$ 0.39 & 24.18 $\pm$ 0.39 \\ \hline 
  $\pi$-$\pi \pi^0$   & 58.39 $\pm$ 0.19 & 21.44 $\pm$ 0.18 \\ \hline 
  $\pi$-$\pi 2 \pi^0$ & 50.36 $\pm$ 0.31 & 34.09 $\pm$ 0.34 \\ \hline 
  $\pi$-$3\pi$        & 54.29 $\pm$ 0.31 & 16.42 $\pm$ 0.28 \\ \hline \hline
%
  $\pi \pi^0$-$\pi \pi^0$   & 59.76 $\pm$ 0.19 & 19.47 $\pm$ 0.17 \\ \hline 
  $\pi \pi^0$-$\pi 2 \pi^0$ & 52.12 $\pm$ 0.22 & 31.92 $\pm$ 0.23 \\ \hline 
  $\pi \pi^0$-$3\pi$        & 54.66 $\pm$ 0.21 & 13.96 $\pm$ 0.19 \\ \hline \hline
%
  $\pi 2 \pi^0$-$\pi 2 \pi^0$ & 45.84 $\pm$ 0.50 & 42.73 $\pm$ 0.56 \\ \hline 
  $\pi 2 \pi^0$-$3\pi$        & 46.98 $\pm$ 0.35 & 27.69 $\pm$ 0.39 \\ \hline \hline
%
  $3\pi$-$3\pi$        & 50.98 $\pm$ 0.48 & 8.57 $\pm$ 0.36 \\ \hline \hline
\end{tabular}
\end{center}
\end{table}

\subsection{Tau direction of flight}

The reconstruction of the $\tau$ flight direction is mandatory in this
analysis to access the event observables, which are functions of 
the four-momenta of the final state particles in the $\tau$ rest frame. 
This can be achieved in the semileptonic decays, for which
the $\tau$ direction lies on a cone around the total
hadron momentum.
For events with both taus decaying semileptonically
the $\tau$ direction 
lies along one of the intersection lines of the two reconstructed cones
in the case that the taus are produced
back-to-back, with equal energies given by $\sqrt{s}/2$, and 
$m_{\nu_{\tau}}= m_{\bar{\nu}_{\tau}} = 0 $.
However, the two cones may not intersect 
due to detector effects or radiation.
If the cones intersect,
the event is considered twice using either solution.
If the cones do not intersect, 
the particle momenta
are fluctuated within their measurement errors and
the event is accepted if the cones intersect in a minimum number of
trials; the average direction is then used~\cite{aleph_polarisation_paper}.

The effect of using both $\tau$ directions has been studied
with a Monte Carlo sample having approximately the same size as the data.
Table~\ref{t_study_cal} presents the statistical errors obtained
in this analysis and using the correct $\tau$ direction from the information
at the generator level. Not distinguishing between the two
$\tau$ directions induces some degradation in 
the overall sensitivity for 
the four anomalous weak dipole moments.
\begin{table}[tbhp]
\caption {\protect\footnotesize Statistical errors obtained from a Monte Carlo
sample approximately equal in size to the data sample, using this analysis
and selecting the correct $\tau$ direction from the 
information at the generator level.
}
\label{t_study_cal}
\begin{center}
\begin{tabular}{|c|c|c|} \hline
  & This analysis & Correct $\tau$ dir.  \\
\hline
$\sigma_{{\rm Re}\, \mu_{\tau} }\,[10^{-3} ]$ & 0.43 & 0.34  \\
$\sigma_{{\rm Im}\,  \mu_{\tau} }\,[10^{-3} ]$ & 0.76 & 0.58 \\
\hline
$\sigma_{{\rm Re}\, d_{\tau} }\,[10^{-3} ]$  & 0.39  & 0.36 \\
$\sigma_{{\rm Im}\, d_{\tau} }\,[10^{-3} ]$  & 0.65  & 0.55 \\
\hline
\end{tabular}
\end{center}
\end{table}

\subsection{Candidates and efficiency matrix}

The final selection for this analysis requires the $\tau$ direction to
be successfully reconstructed, as described in Section~4.2, and the
event observables ($W_1$, $\cos {\theta}_{h_1}$, $\phi_{h_1}$, 
$W_2$, $\cos {\theta}_{h_2}$, $\phi_{h_2}$) to lie in their 
domains of validity.
These requirements decrease the number of candidates by 21\%,
the main reason being the inability to reconstruct the
$\tau$ direction for some events. 

\begin{figure}[htbp]
\begin{center}
\mbox{\epsfig{file=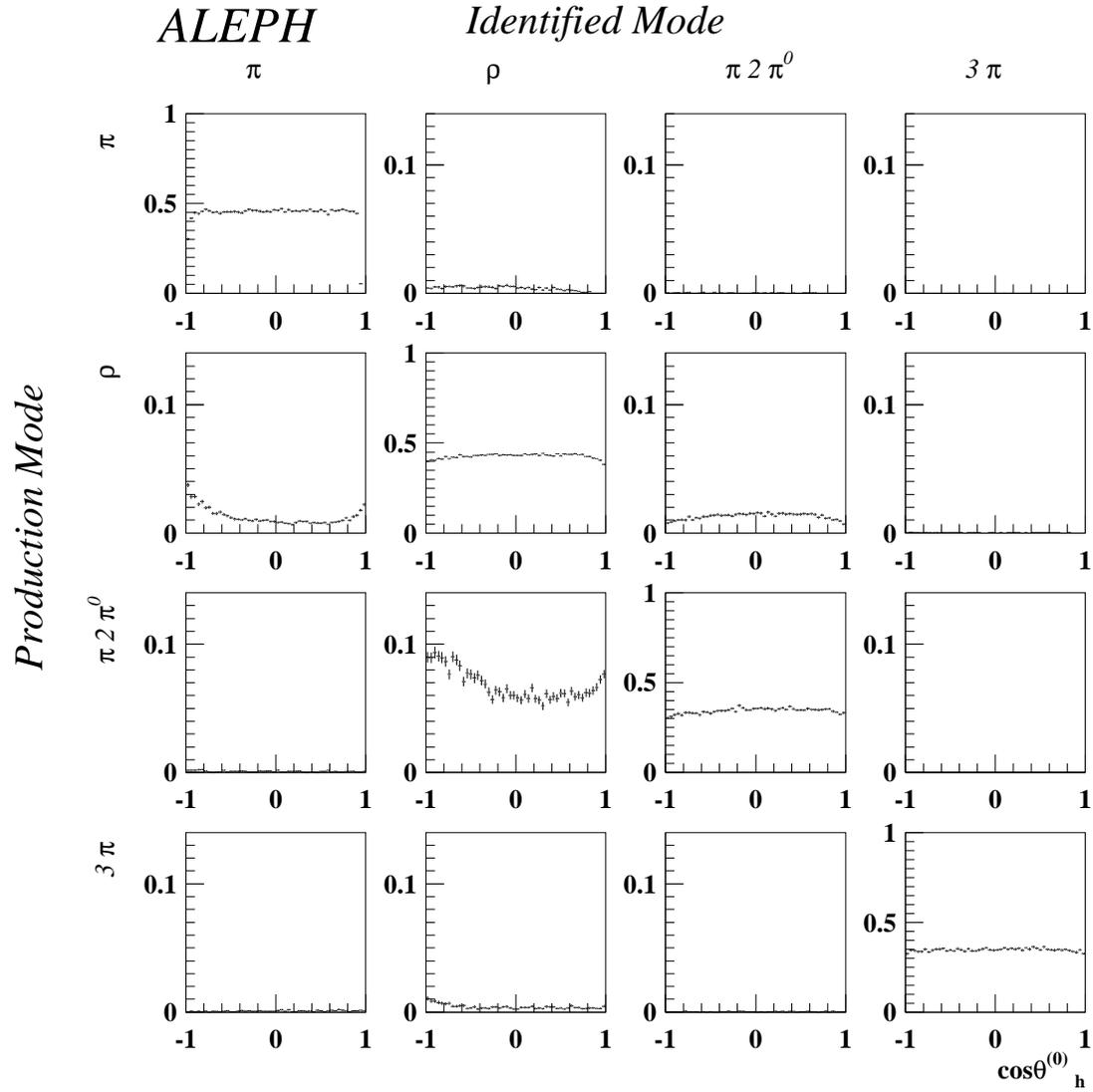,width=16.cm}}
\caption[\protect\footnotesize Efficiency matrix in the barrel
]
{\protect\footnotesize
Efficiency function $\epsilon_{ij}(\cos \theta^{(0)}_{h})$ 
in the barrel region. 
}
\label{fig_det_eff_bar}
\end{center}
\end{figure}

\setcounter{footnote}{0}
The final number of candidates 
in each decay topology is given in Table~3. 
The efficiency matrix $\epsilon_{ij}$ for $i,j = \pi$, $\rho$, $\pi 2 \pi^0$, $3\pi$ 
is calculated as a function of the generated polar angle
$\cos \theta^{(0)}_{h}$ separately
in the barrel and endcaps;
the dependence of $\epsilon_{ij}$ on $\phi^{(0)}_{h}$ and $W^{(0)}$ is quite uniform
and has been integrated out.   
Figure~\ref{fig_det_eff_bar} shows the efficiencies for
the barrel. The diagonal
elements of this figure represent the identification 
of each decay mode, while the off-diagonal 
elements represent its misidentification.
\begin{table}[tbhp]
\caption{\protect\footnotesize
Number of final candidates in each decay topology and total number of
events used in the analysis.}
\label{t_candidates}
\begin{center}
\begin{tabular}{|c|c|c|c|} \hline 
Class & Events & Class & Events \\ \hline \hline
  $\pi$-$\pi$                    & 1901 &   $\pi \pi^0$-$\pi 2 \pi^0$   & 6395 \\ \hline 
  $\pi$-$\pi \pi^0$              & 7844 &   $\pi \pi^0$-$\pi 3\pi$        & 5242 \\ \hline
  $\pi$-$\pi 2 \pi^0$            & 2673 &   $\pi 2 \pi^0$-$\pi 2 \pi^0$ & 1125 \\ \hline
  $\pi$-$3\pi$                   & 2040 &   $\pi 2 \pi^0$-$3\pi$        & 1950 \\ \hline
  $\pi \pi^0$-$\pi \pi^0$        & 8624 &   $3\pi$-$3\pi$               &  712 \\ \hline 
\end{tabular}
\end{center}
\begin{center}
\begin{tabular}{|cccc|} \hline
Total number of events: 38506 & & & \\ \hline
\end{tabular}
\end{center}
\end{table}

\subsection{Detector effects}

The correct approach to introduce the detector effects in 
the likelihood formula would be via a smearing function
$T_{ij}$ depending on 12 variables:
the set of event observables and the
corresponding generated values. The indices 
$i$ and $j$ indicate the generated and the reconstructed channel,
respectively. With the notation used here,
$T_{ij} = T_{ij}( W_1, \cos {\theta}_{h_1}, \phi_{h_1}, 
W_2, \cos {\theta}_{h_2}, \phi_{h_2}, 
W_1^{(0)}, \cos {\theta}_{h_1}^{(0)}, \phi_{h_1}^{(0)}, 
W_2^{(0)}, \cos {\theta}_{h_2}^{(0)}, \phi_{h_2}^{(0)})$. 
Because this function cannot be easily calculated, 
the detector effects are parametrised by the factorised
smearing functions $D_{ij}(x, x^{(0)})$ 
($x$ = $W$, $\cos \theta_{h}$, $\phi_{h}$ and 
$i,j$ = $\pi$, $\rho$, $\pi 2 \pi^0$, $3\pi$).
For $x$ = $W$, $\cos \theta_{h}$, correlations are
neglected, while for $x$ = $\cos \theta_{h}$, $\phi_{h}$, the
correlations are taken into account.

The functions $D_{ij}(x, x^{(0)})$ give the probability
that for generated $i$ and reconstructed $j$ 
the smearing introduced by the detector is $(x-x^{(0)})$ for
a certain generated $x^{(0)}$ with reconstructed $x$.
From the definition it follows that 
\begin{equation}
\int D_{ij}(x,x^{(0)}) d x =1 \: .
\label{D_norm}
\end{equation}
\noindent
These functions are obtained with the SM Monte Carlo simulation by binning the
$(x,x^{(0)})$ plane. 
The binning has been chosen small enough to correctly convolve
detector effects with the generated distribution.

In the likelihood expression, Eq.~13, 
the detector effects are included by replacing the functions $H_i^{\mu}$
by
\bea\lefteqn{
\tilde{H}_j^{\mu}(W,\cos {\theta}_{h}, \phi_{h}) = 
\sum_{i} 
\int H_i^{\mu}(W^{(0)},\cos {\theta}^{(0)}_{h}, \phi^{(0)}_{h})  
D_{ij}(W,W^{(0)}) } \\ \nonumber
&&\rule{30mm}{0pt}
\times D_{ij}(\cos {\theta}_h,\cos {\theta}^{(0)}_h) 
D_{ij}(\phi_h,\phi^{(0)}_{h},\cos {\theta}_h)
\epsilon_{ij}(\cos \theta^{(0)}_{h})     d W^{(0)} d\cos {\theta}^{(0)}_{h} d\phi^{(0)}_{h}
\:.\qquad
\label{eq_ef_sum}
\eea
\noindent
The sum runs over all modes $i$ which have been reconstructed as one of the
modes $j$ used in the analysis, whereby all possible $\tau$ decay modes 
are included in $i$.
The $\tau$ branching fractions are taken into account
implicitly in
Eq.~16 because the 
full differential cross section (Eq.~\ref{eq_final_xsect}) contains
the probability of generating a $\tau^+ \tau^-$ pair decaying
into specific decay modes with certain final state topologies.

In terms of the effective functions $\tilde{H}_i^{\mu}$ the likelihood for each
event reads
\begin{equation}
L_{ij} =
\sum_{\mu,\nu =0,\ldots,3} \bar{R}_{\mu \nu}(\mu_{\tau},d_{\tau} | \theta_{\tau})  
\tilde{H}_i^{\mu}(
W_1,\cos {\theta}_{h_1}, \phi_{h_1} )
\tilde{H}_j^{\nu}(
W_2,\cos {\theta}_{h_2}, \phi_{h_2} ) \:.
\end{equation}

\subsection{Calibration curves}

There are two sources for possible bias in the fitting procedure:
($i$) the detector effects are handled by the factorised
smearing functions $D_{ij}$ above which take correlations into
account only partially, and ($ii$) radiative
corrections are not included in the likelihood.
It is thus necessary to evaluate the adequacy of the fitting process.
This is done with the SCOT Monte Carlo program~\cite{scot}, interfaced
with TAUOLA for the $\tau$ decays and
with the full detector simulation. 
The program SCOT describes
$e^+ e^- \rightarrow \tau^+ \tau^-$ production at an energy around the $Z$ peak at tree
level. It includes the anomalous weak dipole moments and
all $\tau$ spin effects. The initial
state radiation is included by adding
a simple radiator function~\cite{radiator}.
 
The checks are performed by generating various
Monte Carlo samples with different values of the anomalous weak dipole moments.
The couplings $a_e$, $v_e$, $a_{\tau}$, $v_{\tau}$ are set to
their SM values.
The anomalous weak dipole moments
${\rm Re}\,\mu_{\tau}$, ${\rm Im}\,\mu_{\tau}$, ${\rm Re}\,d_{\tau}$, ${\rm Im}\,d_{\tau}$ 
are varied one by one in an adequate region around zero. 
The dependence of the reconstructed values on the generated parameters is
taken as linear. 
Significant deviations 
of the slopes from unity are found for certain decay topologies, and the offset
for ${\rm Re}\,\mu_{\tau}$ is not consistent with zero for certain channels.
These effects have been studied and are mostly related to not using the
correct $\tau$ direction and to background effects.
In this analysis, this calibration for each anomalous
weak dipole moment and decay topology is taken into account
to obtain the corresponding individual measurements.
The slopes, offsets and $\chi^2$ of the linear fits 
are presented in Table~4.
\begin{table}[tbhp]
\caption{\protect\footnotesize
Results of the fit of the calibration curves for each
of the decay topologies obtained with the
SCOT program and a first order radiator, for $\mu_{\tau}$
(top) and for $d_{\tau}$ (bottom).
The slope $a$, the offset $b$ and the  $\chi^2$ of the linear fit
are given for every case. The number of degrees of freedom 
is three for these linear fits.
}
\label{t_res_cali}
\begin{center}
\begin{tabular}{|c|ccr|ccr|} \hline
Channel & 
\multicolumn{3}{|c|} {${\rm Re}\,\mu_{\tau}$}& 
\multicolumn{3}{|c|} {${\rm Im}\,\mu_{\tau}$} \\ \hline \hline
& $a\,[10^{-2}]$ &$b\,[10^{-4}]$ & $\chi^2$ & $a\,[10^{-2}]$ & $b\,[10^{-4}]$& $\chi^2$ \\ \hline
$\pi$-$\pi$                 & 76.2 $\pm$ 5.9  & 14.1 $\pm$ 3.8  & 3.35 
                             & 92.0 $\pm$ 6.3  & $-$1.7 $\pm$ 4.3 & 0.39 \\ \hline
$\pi$-$\rho$                & 93.3 $\pm$ 2.9  & 12.3 $\pm$ 1.7 & 2.28 
                             & 68.2 $\pm$ 3.6  & 0.1 $\pm$ 2.2 & 5.92  \\ \hline
$\pi$-$\pi 2 \pi^0$         & 104.7 $\pm$ 5.3  & 10.2 $\pm$ 3.2 & 0.09 
                             & 82.8 $\pm$ 6.7  & 0.5 $\pm$ 4.1 & 1.63  \\ \hline
$\pi$-$3 \pi$               & 108.9 $\pm$ 5.4  & 5.4 $\pm$ 3.2  & 0.26 
                             & 77.3 $\pm$ 6.8   & $-$0.4 $\pm$ 4.2 & 2.28  \\ \hline
$\rho$-$\rho$               & 99.1 $\pm$ 3.1  & 9.3 $\pm$ 1.8  & 3.64 
                             & 60.4 $\pm$ 3.9  & $-$1.1 $\pm$ 2.3 & 0.84  \\ \hline
$\rho$-$\pi 2 \pi^0$        & 99.0 $\pm$ 3.9  & 0.1 $\pm$ 2.3  & 0.44 
                             & 69.8 $\pm$ 5.0   & 0.8 $\pm$ 3.0 & 4.45  \\ \hline
$\rho$-$3\pi$               & 89.7 $\pm$ 3.8   & 0.3 $\pm$ 2.2 & 3.55 
                             & 60.0 $\pm$ 4.9  & 0.9 $\pm$ 2.8 & 4.14  \\ \hline
$\pi 2 \pi^0$-$\pi 2 \pi^0$ & 96.9 $\pm$ 9.9  & $-$3.1 $\pm$ 5.7 & 1.72 
                             & 60 $\pm$ 14   & $-$6.1 $\pm$ 7.9 & 0.28  \\ \hline
$\pi 2 \pi^0$-$3\pi$        & 89.5 $\pm$ 6.6   & $-$0.3 $\pm$ 3.9 & 1.75 
                             & 73.4 $\pm$ 8.9   & $-$1.3 $\pm$ 5.3 & 1.10  \\ \hline
$3\pi$-$3\pi$               & 102.0 $\pm$ 9.6  & $-$9.6 $\pm$ 5.7 & 2.00 
                             & 55.0 $\pm$ 12.0  & $-$4.0 $\pm$ 7.3 & 1.10  \\ \hline \hline
Channel & 
\multicolumn{3}{|c|} {${\rm Re}\,d_{\tau}$}& 
\multicolumn{3}{|c|} {${\rm Im}\,d_{\tau}$} \\ \hline \hline
& $a\,[10^{-2}]$&$b\,[10^{-4}]$&$\chi^2$&$a\,[10^{-2}]$ &$b\,[10^{-4}]$ & $\chi^2$ \\ \hline
$\pi$-$\pi$                 & 76.2 $\pm$ 4.1 & $-$1.6 $\pm$ 2.6  & 0.25 
                             & 51.3 $\pm$ 5.1  & 5.7 $\pm$ 3.1 & 0.26 \\ \hline
$\pi$-$\rho$                & 109.4 $\pm$ 2.9  & 1.2 $\pm$ 1.8 & 5.82 
                             & 90.5 $\pm$ 3.6  & 4.1 $\pm$ 2.3 & 4.31  \\ \hline
$\pi$-$\pi 2 \pi^0$         & 106.9 $\pm$ 5.4  & 1.3 $\pm$ 3.3 & 5.33 
                             & 91.4 $\pm$ 7.0  & 4.3 $\pm$ 4.4 & 1.17  \\ \hline
$\pi$-$3 \pi$               & 113.5 $\pm$ 5.3  & 6.9 $\pm$ 3.3 & 1.27 
                             & 101.9 $\pm$ 7.0  & 2.2 $\pm$ 4.2 & 4.20  \\ \hline
$\rho$-$\rho$               & 109.1 $\pm$ 3.2  & $-$3.3 $\pm$ 1.9 & 1.83 
                             & 80.6 $\pm$ 4.1  & 0.6 $\pm$ 2.4 & 2.14  \\ \hline
$\rho$-$\pi 2 \pi^0$        & 102.7 $\pm$ 4.0  & $-$1.1 $\pm$ 2.3 & 1.46 
                             & 70.4 $\pm$ 5.2  & 3.0 $\pm$ 3.1 & 3.08  \\ \hline
$\rho$-$3\pi$               & 96.4 $\pm$ 3.9  & 1.3 $\pm$ 2.3 & 3.76 
                             & 60.5 $\pm$ 5.0  & 3.3 $\pm$ 2.9 & 1.63  \\ \hline
$\pi 2 \pi^0$-$\pi 2 \pi^0$ & 96.1 $\pm$ 9.9  & 0.5 $\pm$ 5.8 & 0.44 
                             & 75 $\pm$ 13  & 2.9 $\pm$ 7.8 & 1.08  \\ \hline
$\pi 2 \pi^0$-$3\pi$        & 85.2 $\pm$ 6.8  & 4.3 $\pm$ 4.0 & 1.59 
                             & 66.4 $\pm$ 9.2  & $-$1.1 $\pm$ 5.3 & 0.16  \\ \hline
$3\pi$-$3\pi$               & 97.0 $\pm$ 10.0  & 2.0 $\pm$ 5.8 & 2.48 
                             & 60.0 $\pm$ 12.0  & $-$2.1 $\pm$ 7.3 & 1.73  \\ \hline
\end{tabular}
\end{center}
\end{table}

The offsets and slopes were derived with a Monte Carlo
program which includes only a first order radiator for the initial state
bremsstrahlung. The offsets have then been verified
with the KORALZ Monte Carlo program~\cite{KORALZ} interfaced with
the full detector simulation. KORALZ describes 
$e^+ e^- \rightarrow \tau^+ \tau^-$ 
production at an energy around the $Z$ peak and with initial state
bremsstrahlung corrections up to $O(\alpha^2)$, final state
$O(\alpha)$ bremsstrahlung and $O(\alpha)$ electroweak corrections.
However, this program contains only longitudinal spin effects 
(i.e., the production terms used are $R_{00}$, $R_{03}$ and $R_{33}$) 
and the anomalous weak dipole moments are set to zero. 
To check the effect of this approximation, a maximum likelihood fit was
built to include the complete $R_{00}$, $R_{03}$ and $R_{33}$ terms and
the anomalous parts of the other $R_{\mu \nu}$ terms. The values obtained for
the four anomalous weak dipole moments for each decay topology
are consistent with the
corresponding offsets computed with SCOT and the first order radiator.
Table~\ref{koralz_correction} shows the global differences between the offsets of the
central values of the four anomalous weak dipole moments when using
either KORALZ or SCOT.
The final results of this analysis are obtained by correcting 
the individual measurements
according to the KORALZ offsets and the SCOT slopes and by 
including the statistical error of the correction
in the systematic uncertainty. 
The statistical error of the SCOT
offsets is also included in the systematic uncertainty
as these offsets have contributions from all $R_{\mu \nu}$ terms.
\begin{table}[tbhp]
\caption{\protect\footnotesize
Global differences in the
central values of the four anomalous weak dipole moments when using
offsets from the KORALZ or SCOT programs. 
}
\label{koralz_correction}
\begin{center}
\begin{tabular}{|c|c|} \hline
Parameter & Shift   \\
\hline
${\rm Re}\, \mu_{\tau} \,[10^{-3} ]$ & $-$0.45 $\pm$ 0.21  \\
${\rm Im}\,  \mu_{\tau} \,[10^{-3} ]$ & $-$0.43 $\pm$ 0.42 \\
\hline
${\rm Re}\, d_{\tau} \,[10^{-3} ] $  & 0.05   $\pm$ 0.19   \\
${\rm Im}\, d_{\tau} \,[10^{-3} ]$  & 0.81   $\pm$ 0.38   \\
\hline
\end{tabular}
\end{center}
\end{table}

\section{Systematic uncertainties}
 \label{sys_errors}

The systematic uncertainty is calculated for each anomalous weak dipole 
moment and final state decay topology. The estimates are shown in 
Tables 6 to 9. The last row of these tables
shows the combined systematic uncertainty from each source taking
into account the correlations between channels.

\begin{table}[tbhp]
\caption[\protect\footnotesize Systematic uncertainties on ${\rm Re}\, \mu_{\tau}$
]
{\protect\footnotesize
Systematic uncertainties on ${\rm Re}\, \mu_{\tau}$ for
the different channels.
The last row gives 
the combined systematic uncertainty from
each source taking into account the correlations
between channels.
The total systematic and statistical errors 
are shown in the last two columns.
The values are expressed in units of $[10^{-4}]$.
The sources of uncertainty are explained in the text.
}
\label{t_sreat}
\begin{center}
\begin{tabular}{|c|c|c|c|c|c|c|c|c|c|c|} \hline
& ECAL & TPC & Align.&$\tau$BF&Wpar&MC st. & $a_1$ dyn.&Fake $\gamma$ & $\sigma_{\rm sys}$ & $\sigma_{\rm stat}$ 
 \\  \hline 
$\pi$-$\pi$                  & 4.53 &0.84 &0.58  & 1.55   & 0.11   & 14.30  & 0.& 2.88 & 15.39 &24.70 
\\  \hline 
$\pi$-$\rho$                 & 2.90 &0.48&0.04  & 0.25   & 0.12   & 4.64  & 0. & 0.23 & 5.51  & 9.35
\\  \hline 
$\pi$-$\pi 2 \pi^0$          & 2.77 &2.98&0.12  & 1.12   & 0.06   & 7.52  & 0.45 & 0.54 & 8.65 &14.81
\\  \hline 
$\pi$-$3 \pi$                & 0.63 &12.23&6.95  & 0.43   & 0.14   & 7.39  & 1.23 &0.85 & 15.98&16.58 
\\  \hline 
$\rho$-$\rho$                & 2.63 &1.23& 0.28 & 0.37   & 0.11   & 4.43  & 0.  & 0.64 & 5.36 &8.45
\\  \hline 
$\rho$-$\pi 2 \pi^0$         & 3.20 &0.53&0.42  & 0.61   & 0.06   & 5.53  & 1.88 & 1.54 & 6.90 &11.67
\\  \hline 
$\rho$-$3\pi$                & 1.55 &1.38&  1.94  & 0.60   & 0.08   & 5.87  & 0.61 & 1.60 &6.77&11.37 
\\  \hline 
$\pi 2 \pi^0$-$\pi 2 \pi^0$  & 10.24& 1.72& 0.86  & 1.49   & 0.14   & 13.56  & 11.37&2.03&20.69&22.75 
\\  \hline 
$\pi 2 \pi^0$-$3\pi$         & 4.50 &3.04& 2.38  & 1.62   & 0.03   & 10.07  & 2.91 & 1.31& 12.22 &20.31
 \\  \hline 
$3\pi$-$3\pi$                & 1.65 &7.82&7.62  & 0.89   & 0.24   & 13.17  & 4.69 & 4.61 & 18.43&25.55
\\  \hline 
\hline
  Combined         &0.72 &0.92 &0.43 &0.17 &0.08 &2.19 &0.43 &0.18 &2.60 & 4.20
\\  \hline 
\end{tabular}
\end{center}
\end{table}
\begin{table}[tbhp]
\caption[\protect\footnotesize Systematic uncertainties on ${\rm Im}\, \mu_{\tau} $
]
{\protect\footnotesize
Systematic uncertainties on ${\rm Im}\, \mu_{\tau}$ for
the different channels.
The last row gives 
the combined systematic uncertainty from
each source taking into account the correlations
between channels.
The total systematic and statistical errors 
are shown in the last two columns.
The values are expressed in units of $[10^{-4}]$.
The sources of uncertainty are explained in the text.
}
\label{t_simat}
\begin{center}
\begin{tabular}{|c|c|c|c|c|c|c|c|c|c|c|} \hline
& ECAL & TPC & Align. &$\tau$BF &Wpar & MC st. & $a_1$ dyn.&Fake $\gamma$ & $\sigma_{\rm sys}$ & $\sigma_{\rm stat}$ 
 \\  \hline 
$\pi$-$\pi$                  & 1.60 &0.86&0.42& 0.43 & 0.09 & 13.29  & 0. & 0.97& 13.46&25.06 
\\  \hline 
$\pi$-$\rho$                 & 0.86 &0.38& 1.52 & 0.39 & 0.05 & 8.11& 0. & 2.44& 8.66&16.60   
\\  \hline 
$\pi$-$\pi 2 \pi^0$          & 27.55 &3.40&3.03& 1.58 & 0.39 & 14.92& 8.75 & 3.24 &33.05&29.65   
\\  \hline 
$\pi$-$3 \pi$                & 11.23 &9.64&9.98& 0.61 & 0.27 & 15.72& 8.94 & 0.& 25.42&32.96   
\\  \hline 
$\rho$-$\rho$                & 3.29 &0.13&0.85& 0.64 & 0.09 & 10.21& 0. & 1.09 & 10.84&18.09   
\\  \hline  
$\rho$-$\pi 2 \pi^0$         & 32.05 &2.98&1.13& 0.56 & 0.06 & 12.51& 2.01 & 8.19 & 35.57&25.10   
\\  \hline 
$\rho$-$3\pi$              & 9.60 &1.51& 0.57& 1.43 & 0.08 & 12.70& 2.04 & 10.85 & 19.40&21.85   
\\  \hline 
$\pi 2 \pi^0$-$\pi 2 \pi^0$  & 100.21 &2.20& 6.70 & 5.49 & 0.57 & 35.31& 7.17 & 10.17&107.35&62.38   
\\  \hline 
$\pi 2 \pi^0$-$3\pi$         & 17.50 &11.90& 5.75 & 1.14 & 0.22 & 17.61& 12.21 & 15.62 & 34.43&37.88   
 \\  \hline 
$3\pi$-$3\pi$                & 5.38 &5.70& 1.66& 0.99 & 0.31 & 34.50& 10.76 & 17.05& 40.77&64.46   
\\  \hline
\hline
 Combined          &4.27 &0.17 &0.30 &0.23 &0.05 &4.18 &1.12 &0.76 &6.10 & 8.00
\\  \hline 
\end{tabular}
\end{center}
\end{table}

The ECAL effects are related to the uncertainty in the global energy scale and 
the nonlinearity of the response.
The global scale is known at the level of 0.25\%, through the calibration 
with Bhabha events. 
Global variations have been applied to each ECAL module and the
effect is propagated to the fitted parameters. 
The nonlinearity of the response is related to the wire 
saturation constants of ECAL, which are different in the
barrel and endcaps. These saturation constants have been fluctuated 
within their nominal errors while
keeping the measured energy fixed at $M_{\rm Z} /2$.

The TPC systematic errors are 
related to the momentum measurement:
$(i)$ an effect due to the magnetic field
acting similarly on positive and negative charged tracks, 
and $(ii)$ a sagitta effect affecting oppositely positive 
and negative tracks. These two effects are calibrated 
with dimuon events and the corresponding
corrections are applied to the $\tau$ data. The systematic
errors are then estimated by varying 
the corrections within their errors for each year.

The systematic errors in the column labeled ``Align.'' are due to
a possible azimuthal tilt between the different parts of the detector.

Variations in the $\tau$ branching fractions are considered in
the $\tau$BF column. 
This systematic uncertainty was determined from its effect
on the calibration curves (Section~4.5).

The experimental errors on the weak parameters $\sin^2 \theta_W$ and $M_{\rm Z}$
are propagated to the fitted values. Other 
weak parameters have negligible effect on the measurements.
These effects are summarised in the column labeled
``Wpar''.

The finite Monte Carlo statistics also causes systematic uncertainties.
The most relevant statistical uncertainty is for 
the KORALZ offsets.
The statistical error of the offsets and slopes obtained with
SCOT and the first order radiator (Table~\ref{t_res_cali})
are also taken into account.
Finally,
the statistical error in the calculation of the efficiency matrix is
also considered.
These effects are shown under the column ``MC st.''.

\begin{table}[tbhp]
\caption[\protect\footnotesize Systematic uncertainties on ${\rm Re}\, d_{\tau}$
]
{\protect\footnotesize
Systematic uncertainties on ${\rm Re}\, d_{\tau}$ for
the different channels.
The last row gives 
the combined systematic uncertainty from
each source taking into account the correlations
between channels.
The total systematic and statistical errors 
are shown in the last two columns.
The values are expressed in units of $[10^{-4}]$.
The sources of uncertainty are explained in the text.
}
\label{t_sredt}
\begin{center}
\begin{tabular}{|c|c|c|c|c|c|c|c|c|c|c|} \hline
& ECAL & TPC & Align. &$\tau$BF & Wpar & MC st. & $a_1$ dyn.&Fake $\gamma$ & $\sigma_{\rm sys}$ & $\sigma_{\rm stat}$ 
 \\  \hline 
$\pi$-$\pi$                  & 1.27 &0.58&0.09& 0.77 & 0.12 & 8.33& 0. & 0.16& 8.48&15.61   
\\  \hline 
$\pi$-$\rho$                 & 0.18 &0.40&0.61& 0.35 & 0.04 & 3.91& 0. & 4.51 & 6.03&7.94   
\\  \hline 
$\pi$-$\pi 2 \pi^0$        & 8.38 &1.73&1.58  & 1.78 & 0.09 & 7.68& 3.36 & 3.21& 12.63&15.34   
\\  \hline 
$\pi$-$3 \pi$                & 1.36 &0.57&0.02& 0.27 & 0.02 & 7.22& 2.71 & 1.00 & 7.92&15.05   
\\  \hline 
$\rho$-$\rho$                & 0.53 &0.35&0.49& 0.38 & 0.04 & 4.18& 0. & 0.17& 4.28&8.04   
\\  \hline 
$\rho$-$\pi 2 \pi^0$         & 1.45 &1.68&0.63& 0.79 & 0.08 & 5.39& 0.79 & 2.63& 6.52 &10.89  
\\  \hline 
$\rho$-$3\pi$                & 5.37 &1.68&0.79& 0.45 & 0. & 5.56& 3.00 & 2.26 & 8.80& 10.89 
\\  \hline 
$\pi 2 \pi^0$-$\pi 2 \pi^0$& 8.75 &2.00&1.10& 1.71 & 0.07 & 13.35& 8.18 & 9.44 & 20.47&29.74   
\\  \hline 
$\pi 2 \pi^0$-$3\pi$       & 5.26 &12.07&10.96& 2.56 & 0.05 & 10.39& 4.88 & 3.75& 21.11&21.65   
 \\  \hline 
$3\pi$-$3\pi$              & 1.91 &4.75&4.70& 1.64 & 0.12 & 14.06& 5.34 & 2.25& 16.80&29.39   
\\  \hline
\hline
   Combined       &0.41 &0.29 &0.26 &0.17 &0.02 &1.94 &0.90 &0.50 &2.30 & 3.90
\\  \hline 
\end{tabular}
\end{center}
\end{table}
\begin{table}[tbhp]
\caption[\protect\footnotesize Systematic uncertainties on ${\rm Im}\, d_{\tau} $
]
{\protect\footnotesize
Systematic uncertainties on ${\rm Im}\, d_{\tau}$ for
the different channels.
The last row gives 
the combined systematic uncertainty from
each source taking into account the correlations
between channels.
The total systematic and statistical errors 
are shown in the last two columns.
The values are expressed in units of $[10^{-4}]$.
The sources of uncertainty are explained in the text.
}
\label{t_simdt}
\begin{center}
\begin{tabular}{|c|c|c|c|c|c|c|c|c|c|c|} \hline
& ECAL & TPC & Align. &$\tau$BF &Wpar & MC st. & $a_1$ dyn.&Fake $\gamma$ & $\sigma_{\rm sys}$ & $\sigma_{\rm stat}$ 
 \\  \hline 
$\pi$-$\pi$                & 1.02 &0.59&0.65& 1.25 & 0.06 & 14.95& 0. & 5.01 & 15.87& 26.10   
\\  \hline 
$\pi$-$\rho$               & 2.48 &0.74&0.41& 0.16 & 0.08 & 6.90& 0. & 0.94& 7.44&13.63   
\\  \hline 
$\pi$-$\pi 2 \pi^0$        & 51.04 &6.50& 2.73 &1.14& 0.81 & 15.60& 8.59 & 5.29& 54.79&29.55   
\\  \hline 
$\pi$-$3 \pi$              & 3.47 &3.38&5.14& 0.79   & 0.14 & 11.90& 4.30 & 8.03& 16.59&23.50   
\\  \hline 
$\rho$-$\rho$              & 2.91 &0.60&0.55& 0.63   & 0.04 & 8.54& 0. & 0.26 & 9.09&15.07   
\\  \hline 
$\rho$-$\pi 2 \pi^0$       & 17.69 &0.44&0.29& 1.20  & 0.08 & 12.46& 6.52 & 2.54 & 22.78&24.64   
\\  \hline 
$\rho$-$3\pi$              & 6.09 &9.99& 5.87 & 0.36 & 0.07 & 12.40& 1.04 & 5.91& 19.01&23.31   
\\  \hline 
$\pi 2 \pi^0$-$\pi 2 \pi^0$& 9.86 &10.57&17.41& 3.38 & 1.15 & 28.17& 40.10 & 14.34 & 55.96&72.95   
\\  \hline 
$\pi 2 \pi^0$-$3\pi$       & 8.12 &7.21&3.72& 4.21   & 0.11 & 19.32& 1.73 & 6.86 & 23.93&41.51   
 \\  \hline 
$3\pi$-$3\pi$              & 4.46 &12.78&8.80& 0.67  & 0.29 & 30.10& 7.80 & 4.19& 35.30&59.22   
\\  \hline
\hline
     Combined     &5.74 &1.80 &0.70 &0.32 &0.04 &3.80 &0.79 &0.43 &7.20 & 7.20
\\  \hline 
\end{tabular}
\end{center}
\end{table}
The $a_1$ decay dynamics are not well described theoretically.
The impact of this was evaluated in the past~\cite{polari_duflot} by implementing 
several models in the analysis~\cite{L_Duflot_thesis}.
The implementation of those models is much more difficult in the present analysis.
The uncertainty is estimated by means of three models:
the K{\"u}hn \& Santamaria (KS) model~\cite{Santamaria} (used in the fitting formula),
the Feindt model~\cite{Feindt} and the 
Isgur, Morningstar and Reader (simplified) model~\cite{IMR}. The effects of the Feindt
and IMR models on $W$ and $h^3$ are calculated. The corresponding ratios 
with the $W$ and the $h^3$ of the KS model are then used to scale the error. 

Another source of systematic error is due to fake photons
generated by hadron interactions in the ECAL or by electromagnetic fluctuations.
This quantity of photon candidates is underestimated in the Monte Carlo
simulation compared to the data. This deficit was originally 
observed in a substantial disagreement between the data and the Monte Carlo 
simulation
for the $W$ distribution in the $\pi 2 \pi^0$ channel. This discrepancy
has notably decreased
after weighting the events of the Monte Carlo simulation
according to the number of fake photons. This weighting
was optimised for other $\tau$ analyses~\cite{aleph_polarisation_paper}. 
Figure~\ref{f_obs_w} compares the $W$ distributions for the
data and the Monte Carlo after the approximate weighting. 
In the end, the effect of fake photons is taken into account by
removing fake photons in the simulation
and using the difference in the fitted parameters as systematic uncertainty.
%

The total systematic error and the statistical error for each channel and parameter 
are shown in the last two columns of the tables.

\section{Results and conclusions}
 \label{results}

\begin{figure}[htbp]
\begin{center}
\mbox{\epsfig{file=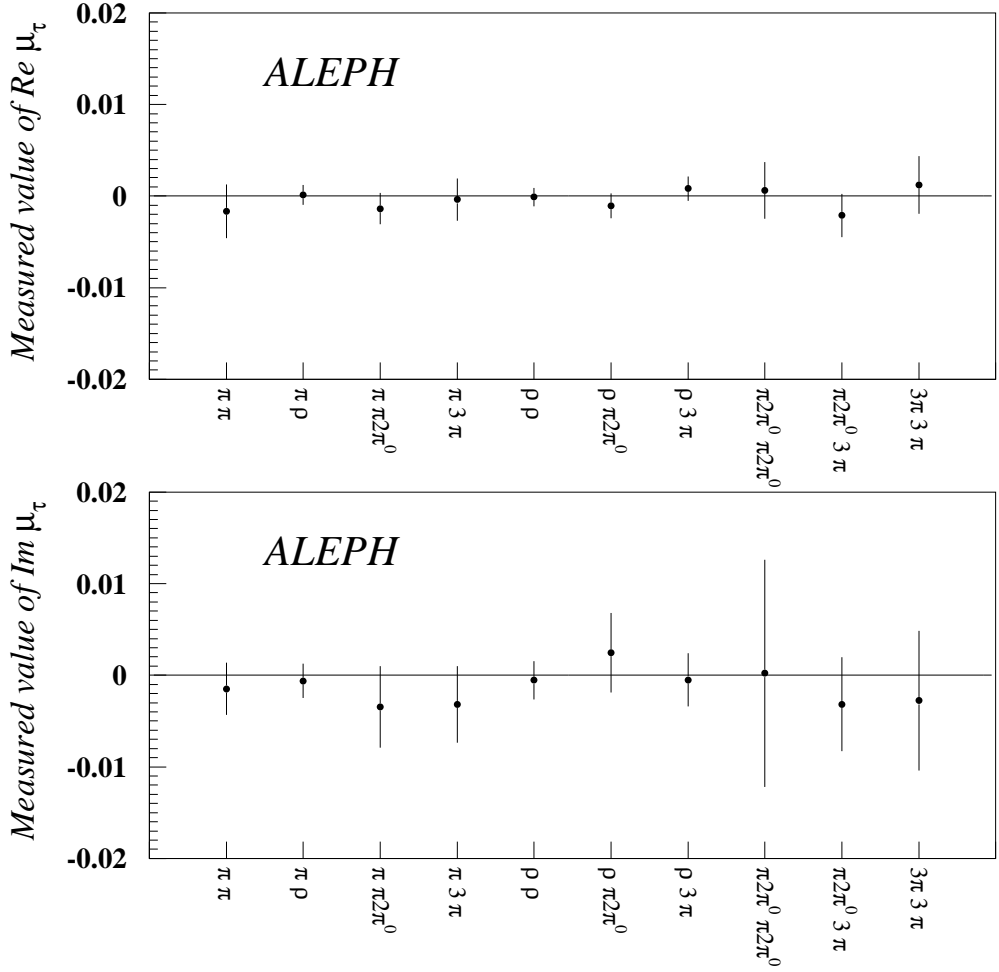,width=15.cm}}
\caption[\protect\footnotesize First results on the anomalous weak
magnetic dipole moment channel by channel
]
{\protect\footnotesize
Results on $\mu_{\tau}$ for the various decay modes, including
both the statistical and the systematic uncertainties. The results on 
${\rm Re}\,\mu_{\tau}$ are shown at the top, and on
${\rm Im}\,\mu_{\tau}$ at the bottom. 
}
\label{fig_resu_1}
\end{center}
\end{figure}

\begin{figure}[htbp]
\begin{center}
\mbox{\epsfig{file=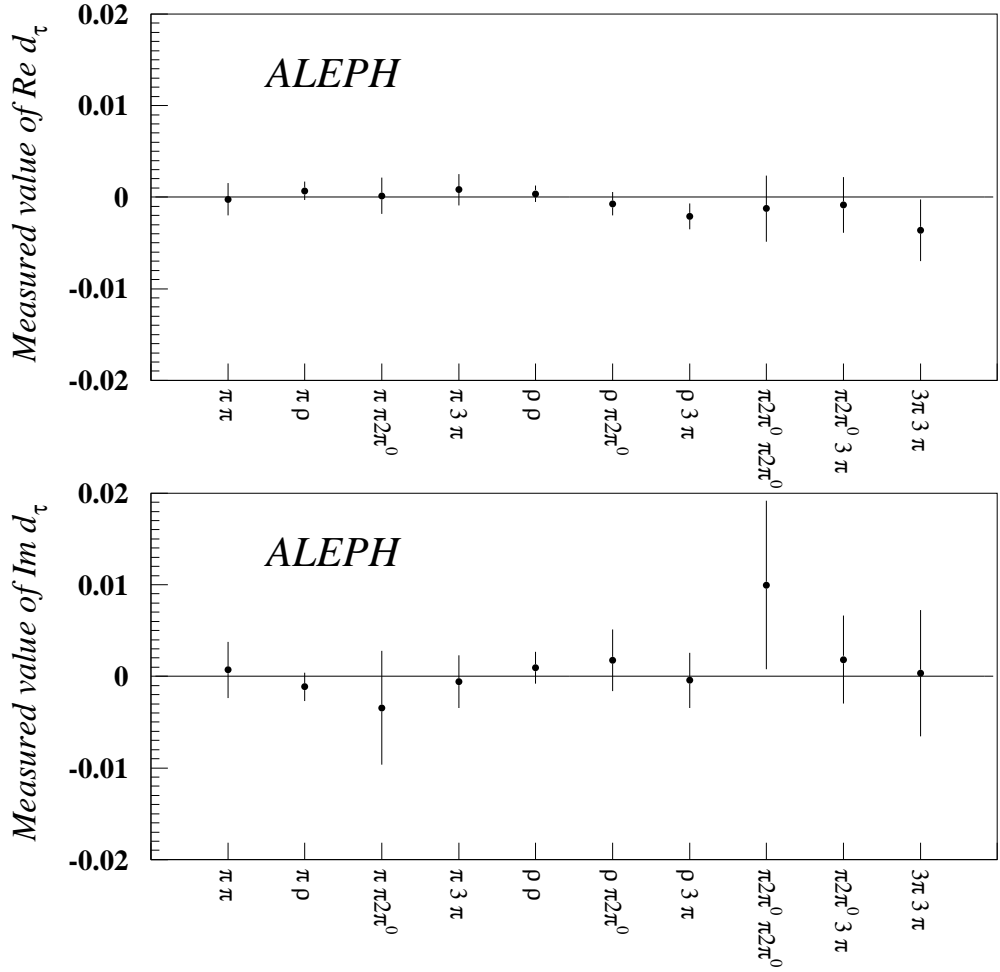,width=15.cm}}
\caption[\protect\footnotesize First results on the anomalous weak
electric dipole moment channel by channel
]
{\protect\footnotesize
Results on $d_{\tau}$ for the various decay modes, including
both the statistical and the systematic uncertainties. The results on 
${\rm Re}\,d_{\tau}$ are shown at the top, and on
${\rm Im}\,d_{\tau}$ at the bottom. The anomalous weak electric dipole
moment is assumed dimensionless in these figures.
}
\label{fig_resu_2}
\end{center}
\end{figure}

\begin{figure}[htbp]
\begin{center}
\mbox{\epsfig{file=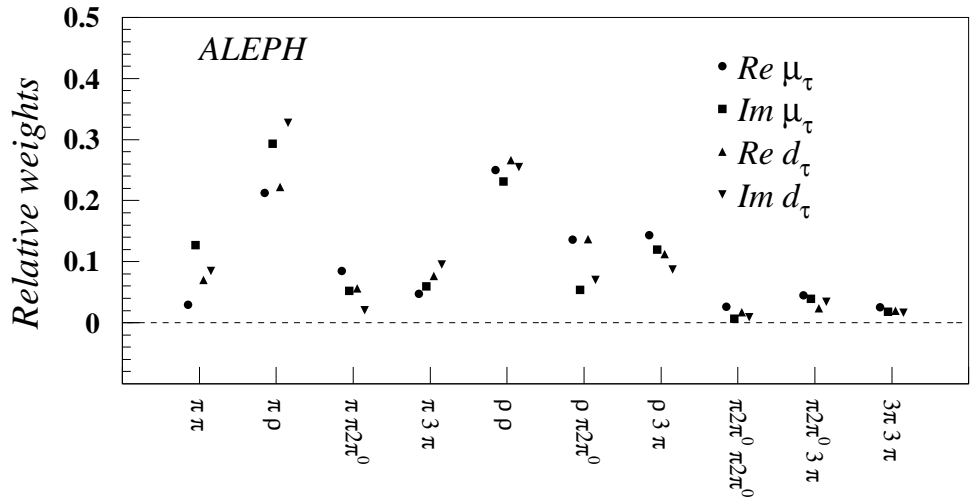,width=15.cm}}
\caption[\protect\footnotesize First results on the anomalous weak
electric dipole moment channel by channel
]
{\protect\footnotesize
Relative weights of the different decay topologies
for the four measured anomalous weak dipole moments, 
normalised to the total weight.
}
\label{fig_weights}
\end{center}
\end{figure}

The final individual measurements of the 
four anomalous weak dipole moments are obtained applying 
the offsets and slopes described in Section~4.5.
The results for the different decay topologies are presented
in Figs.~\ref{fig_resu_1} and~\ref{fig_resu_2}, including both  
the systematic and the statistical errors.
All these measurements are consistent with the SM prediction.
Figure~\ref{fig_weights} shows the relative weights of the different
decay channels for the four measured anomalous weak dipole moments.

The final combined results on the four anomalous weak dipole moments are listed in 
Table~\ref{t_fit_resu_last}, showing the  
statistical, systematic and total errors. The statistical correlations 
are given in Table~\ref{t_fit_corr}.
The final 95\% CL upper limits derived from these measurements are presented
in Table~\ref{t_lim_fin}.
\begin{table}[tbhp]
\caption[\protect\footnotesize Final results at 68 \% CL
]
{\protect\footnotesize
\protect\footnotesize Final results on the real and imaginary terms of 
the anomalous weak dipole moments. 
}
\label{t_fit_resu_last}
\begin{center}
\begin{tabular}{|c|c|c|c|c|} \hline
Parameter & Fitted value & $\sigma_{\rm stat}$ & $\sigma_{\rm sys}$ &$\sigma$\\
\hline
${\rm Re}\, \mu_{\tau}\,[10^{-3}]$ & $-$0.33 & 0.42  & 0.26 &0.49\\
${\rm Im}\,  \mu_{\tau}\,[10^{-3}]$ & $-$0.99 & 0.80 & 0.61 &1.01\\
\hline
${\rm Re}\, d_{\tau}\,[10^{-3} ]\,([10^{-18} {\rm e \, cm}]) $ & $-$0.11 ($-$0.59) & 0.39 (2.14) & 0.23 (1.26) &0.45 (2.49)\\
${\rm Im}\, d_{\tau}\,[10^{-3} ]\,([10^{-18} {\rm e \, cm}]) $ & $-$0.08 ($-$0.45) &0.72 (4.00) & 0.72 (4.01) &1.02 (5.67)\\
\hline
\end{tabular}
\end{center}
\end{table}
\begin{table}[tbhp]
\caption[\protect\footnotesize Correlation matrix between the
fitted parameters
]
{\protect\footnotesize Statistical correlations between the fitted
parameters. The individual correlations are presented in the
off-diagonal elements.
}
\label{t_fit_corr}
\begin{center}
\begin{tabular}{|c|c|c|c|c|} \hline
 &${\rm Re}\, \mu_{\tau}$ & ${\rm Im}\, \mu_{\tau}$ & ${\rm Re}\, d_{\tau}$& ${\rm Im}\, d_{\tau}$ \\
\hline
${\rm Re}\, \mu_{\tau} $ & 1.0 & 0.006 & 0.028 & 0.062 \\
${\rm Im}\, \mu_{\tau} $ & & 1.0 & $-$0.055 & 0.034 \\
\hline
${\rm Re}\, d_{\tau}$  & & & 1.0 & $-$0.003 \\
${\rm Im}\, d_{\tau}$  & & & & 1.0 \\
\hline
\end{tabular}
\end{center}
\end{table}
\begin{table}[tbhp]
\caption
{\protect\footnotesize
\protect\footnotesize Upper limits derived from this measurement of
the anomalous weak dipole moments (95\% CL).
}
\label{t_lim_fin}
\begin{center}
\begin{tabular}{|c|c|} \hline
Parameter & Limit \\
\hline
$|{\rm Re}\, \mu_{\tau}|\,[10^{-3}]$ & 1.14 \\
$|{\rm Im}\, \mu_{\tau}|\,[10^{-3}]$ & 2.65 \\
\hline
$|{\rm Re}\, d_{\tau}|\,[10^{-3} ]\,([10^{-18} {\rm e \, cm}]) $ & 0.91 (5.01)\\
$|{\rm Im}\, d_{\tau}|\,[10^{-3} ]\,([10^{-18} {\rm e \, cm}]) $ & 2.01 (11.15)\\
\hline
\end{tabular}
\end{center}
\end{table}

These results supersede the previous ALEPH measurement of ${\rm Re}\, d_{\tau} $~\cite{dtau_ALEPH};
the measurement of ${\rm Re}\, \mu_{\tau}$, ${\rm Im}\, \mu_{\tau}$
and ${\rm Im}\, d_{\tau}$, presented in this paper gives the
most stringent limits on these quantities to date. 

\section*{Acknowledgments}

We wish to thank our colleagues from the accelerator divisions for the successful
operation of LEP. We are indebted to the engineers and technicians in all our 
institutions for their contribution to the good performance of ALEPH.
Those of us from non-member states thank CERN for its hospitality.

%
%
%
%
%
%

\end{document}